\newif\ifOneCol

\ifOneCol
	\documentclass[draftcls,onecolumn,11pt]{IEEEtran}
\else
	\documentclass[twocolumn,10pt]{IEEEtran}
\fi

\usepackage{color} 
\usepackage[cmex10]{amsmath}
\usepackage{amsfonts}
\usepackage{mathtools} 
\usepackage{verbatim}
\usepackage{multirow}

\usepackage[utf8]{inputenc} 

\usepackage{cite}

\ifCLASSINFOpdf
  \usepackage[pdftex]{graphicx}
  \graphicspath{{graphics/}}
  \DeclareGraphicsExtensions{.pdf,.jpeg,.png}
\else
  \usepackage[dvips]{graphicx}
  \graphicspath{{graphics/}}
  \DeclareGraphicsExtensions{.eps}
\fi

\usepackage{array}

\usepackage{mdwmath}
\usepackage{mdwtab}
\begin{document}
\bibliographystyle{IEEEtran}
%
\title{Joint Channel Parameter Estimation via Diffusive Molecular Communication}

\author{Adam Noel, \IEEEmembership{Student Member, IEEE}, Karen C. Cheung, and
	Robert Schober,	\IEEEmembership{Fellow, IEEE}
\thanks{Manuscript received October 14, 2014; revised April 5, 2015; accepted June 14, 2015. This work was presented in part at IEEE GLOBECOM 2014 \cite{RefWorks:808}. This work was supported by the Natural Sciences and Engineering Research Council of Canada. Computing resources were provided by WestGrid and Compute/Calcul Canada.}
\thanks{The authors are with the Department of Electrical and Computer Engineering, University of British Columbia, Vancouver, BC, Canada (email: \{adamn, kcheung, rschober\}@ece.ubc.ca).}
\thanks{R.~Schober is also with the Institute for Digital Communication, Friedrich-Alexander-Universit\"{a}t Erlangen-N\"{u}rnberg (FAU), Erlangen, Germany (email: schober@lnt.de).}}


\newcommand{\dbydt}[1]{\frac{d#1}{dt}}
\newcommand{\pbypx}[2]{\frac{\partial #1}{\partial #2}}
\newcommand{\psbypxs}[2]{\frac{\partial^2 #1}{\partial {#2}^2}}
\newcommand{\psbypxpy}[3]{\frac{\partial^2 #1}{\partial #2 \partial #3}}
\newcommand{\dbydtc}[1]{\dbydt{\conc{#1}}}
\newcommand{\x}{x}
\newcommand{\y}{y}
\newcommand{\z}{z}
\newcommand{\rad}[1]{\vec{r}_\textnormal{#1}}
\newcommand{\radmag}[1]{|\rad{#1}|}
\newcommand{\vect}[1]{\mathbf{#1}}
\newcommand{\vectm}[1]{\boldsymbol{#1}}


\newcommand{\kth}[1]{k_{#1}}
\newcommand{\Nobs}{{\Nx{\A}}_{\textnormal{ob}}}
\newcommand{\Nobst}[1]{\Nobs\!\left(#1\right)}
\newcommand{\Nxtavg}[2]{\overline{{\Nx{\A}}_\textnormal{#2}}\left(#1\right)}

\newcommand{\M}{M}
\newcommand{\numParam}{L}
\newcommand{\numEst}{Q}
\newcommand{\winLength}{W}
\newcommand{\smM}{m}
\newcommand{\A}{A}
\newcommand{\vx}[1]{v_{#1}}
\newcommand{\vpara}{v_{\scriptscriptstyle\parallel}}
\newcommand{\vperp}{v_{\perp}}
\newcommand{\vxvec}[1]{\mathbf{v}_{#1}}
\newcommand{\metre}{\textnormal{m}}
\newcommand{\second}{\textnormal{s}}
\newcommand{\molecule}{\textnormal{molecule}}
\newcommand{\argmax}{\operatornamewithlimits{argmax}}
\newcommand{\Dx}[1]{D_{#1}}
\newcommand{\Nx}[1]{N_{#1}}
\newcommand{\Nemit}{\Nx{}}
\newcommand{\Vrx}{V_\textnormal{RX}}
\newcommand{\rrx}{r_\textnormal{RX}}
\newcommand{\ttext}[1]{t_\textnormal{#1}}
\newcommand{\stext}[1]{s_\textnormal{#1}}

\newcommand{\textBar}[1]{\big|_\textnormal{#1}}

\newcommand{\EXP}[1]{\exp\left(#1\right)}
\newcommand{\var}[1]{\textnormal{var}(#1)}
\newcommand{\mse}[1]{\textnormal{mse}(#1)}

\newcommand{\w}{w}
\newcommand{\n}{n}
\newcommand{\deltObs}{t_{o}}
\newcommand{\sx}[1]{s_{#1}}

\newtheorem{theorem}{Theorem}
\newtheorem{remark}{Remark}

\newcommand{\edit}[2]{{#1}}

\newcommand{\figModel}[2]{
	\begin{figure}[#1]
		\centering
		\def\svgwidth{#2\linewidth}
		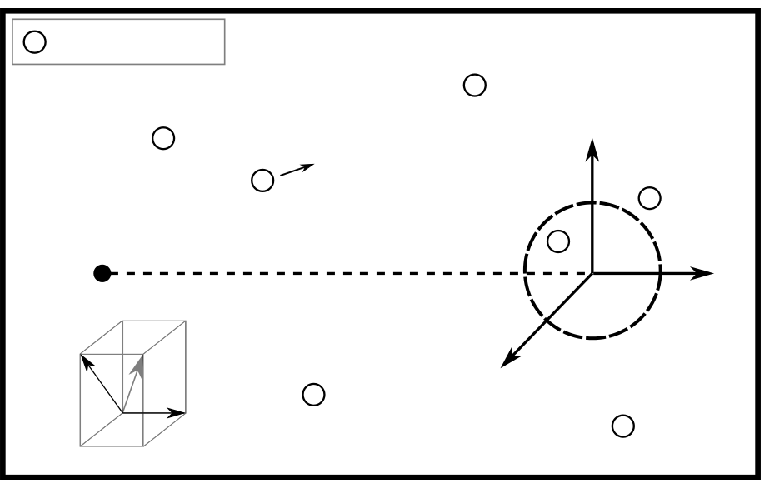
		\caption{The system model considered throughout this paper. The TX is a point source of $\A$ molecules and the RX is a passive observer centered at the origin. The $\A$ molecules are shown as small hallow circles and some are labeled. Molecule $1$ is inside $\Vrx$ and so can be observed by the RX. Molecule $2$ was previously inside $\Vrx$ and is now outside because the RX is non-absorbing. Once released by the TX, the behavior of each molecule is that of a biased random walk (biased by the steady flow $\vxvec{}$) until it undergoes degradation via the chemical reaction described by first-order degradation rate constant $\kth{}$, e.g., molecule $3$.}
		\label{fig_model}
	\end{figure}
}

\newcommand{\tableFIMTwoCol}[1]{
	\begin{table}[#1]
		\centering
		\caption{Diagonal elements of the FIM for each desired parameter.}
		{\renewcommand{\arraystretch}{1.4}
			\begin{tabular}{|p{2.78cm}|c|p{3.72cm}|}
				\hline & & \\[-8pt]
				Parameter & \parbox[c]{1cm}{Variable\\Name $\theta_i$} & FIM Diagonal Element $\left[\mathbf{I}\left(\vectm{\theta}\right)\right]_{\theta_i}$ \\[2mm] \hline & & \\[-11pt]
				Distance from TX to RX  & $d$ & $\sum_{\smM = 1}^\M \frac{\Nxtavg{t_{\smM}}{ob}}{4\Dx{}^2}\left(\vpara-\frac{d}{\ttext{ef}}\right)^2$ \\ \hline & & \\[-11pt]
				TX Release Time & $t_0$	& \parbox[c]{4cm}{$\sum_{\smM = 1}^\M \Nxtavg{t_{\smM}}{ob}\Big(\frac{3}{2\ttext{ef}} + \kth{}$ \\$ + \frac{\vpara^2 + \vperp^2}{4\Dx{}} - \frac{d^2}{4\Dx{}\ttext{ef}^2} \Big)^2$} \\ \hline & & \\[-11pt]
				Diffusion Coefficient & $\Dx{}$ & \parbox[c]{4cm}{$\sum_{\smM = 1}^\M \frac{\Nxtavg{t_{\smM}}{ob}}{4\Dx{}^2}\Big(3$ \\$\quad - \frac{(d - \vpara \ttext{ef})^2 + \vperp^2 \ttext{ef}^2}{2\Dx{}\ttext{ef}}\Big)^2$} \\ \hline & & \\[-11pt]
				Degradation Rate & $\kth{}$ & $\sum_{\smM = 1}^\M \ttext{ef}^2\Nxtavg{t_{\smM}}{ob}$ \\ \hline & & \\[-11pt]
				Flow Towards RX & $\vpara$ & $\sum_{\smM = 1}^\M \frac{\Nxtavg{t_{\smM}}{ob}}{4\Dx{}^2}\left(d-\vpara \ttext{ef}\right)^2$ \\ \hline & & \\[-11pt]
				Perpendicular Flow & $\vperp$ & $\sum_{\smM = 1}^\M \frac{\vperp^2 \ttext{ef}^2}{4\Dx{}^2}\Nxtavg{t_{\smM}}{ob}$ \\ \hline & & \\[-11pt]
				Molecules Released by TX & $\Nemit$ & $\sum_{\smM = 1}^\M \frac{\Nxtavg{t_{\smM}}{ob}}{\Nemit^2}$ \\ \hline
			\end{tabular}
		}
		\label{table_fim}
	\end{table}
}

\newcommand{\tableFIMOneCol}[1]{
	\begin{table}[#1]
		\centering
		\caption{Diagonal elements of the FIM for each desired parameter.}
		{\renewcommand{\arraystretch}{1.4}
			\begin{tabular}{|p{3.5cm}|c|l|}
				\hline
				Parameter & \parbox[c]{1.1cm}{Variable\\Name $\theta_i$} & FIM Diagonal Element $\left[\mathbf{I}\left(\vectm{\theta}\right)\right]_{\theta_i}$ \\[2mm] \hline
				Distance from TX to RX  & $d$ & $\sum_{\smM = 1}^\M \frac{\Nxtavg{t_{\smM}}{ob}}{4\Dx{}^2}\left(\vpara-\frac{d}{\ttext{ef}}\right)^2$ \\ \hline
				TX Release Time & $t_0$	& $\sum_{\smM = 1}^\M \Nxtavg{t_{\smM}}{ob}\left(\frac{3}{2\ttext{ef}} + \kth{} + \frac{\vpara^2 + \vperp^2}{4\Dx{}} - \frac{d^2}{4\Dx{}\ttext{ef}^2} \right)^2$ \\ \hline
				Diffusion Coefficient & $\Dx{}$ & $\sum_{\smM = 1}^\M \frac{\Nxtavg{t_{\smM}}{ob}}{4\Dx{}^2}\left(3 - \frac{(d - \vpara \ttext{ef})^2 + \vperp^2 \ttext{ef}^2}{2\Dx{}\ttext{ef}}\right)^2$ \\ \hline
				Degradation Rate & $\kth{}$ & $\sum_{\smM = 1}^\M \ttext{ef}^2\Nxtavg{t_{\smM}}{ob}$ \\ \hline
				Flow Towards RX & $\vpara$ & $\sum_{\smM = 1}^\M \frac{\Nxtavg{t_{\smM}}{ob}}{4\Dx{}^2}\left(d-\vpara \ttext{ef}\right)^2$ \\ \hline
				Perpendicular Flow & $\vperp$ & $\sum_{\smM = 1}^\M \frac{\vperp^2 \ttext{ef}^2}{4\Dx{}^2}\Nxtavg{t_{\smM}}{ob}$ \\ \hline
				Molecules Released by TX & $\Nemit$ & $\sum_{\smM = 1}^\M \frac{\Nxtavg{t_{\smM}}{ob}}{\Nemit^2}$ \\ \hline
			\end{tabular}
		}
		\label{table_fim}
	\end{table}
}

\newcommand{\tableParam}[1]{
	\begin{table}[#1]
		\centering
		\caption{System parameters used for numerical and simulation results. The ``Min'' and ``Max'' values are the bounds of ML estimation via grid search.}
		{\renewcommand{\arraystretch}{1.4}
			\begin{tabular}{|l|c|c|c|c|c|}
				\hline
				Parameter & Symbol & Units & Value & Min & Max \\ \hline
				RX Radius & $\rrx$	& $\mu\metre$ & $0.5$ & - & -  \\ \hline
				Sim. Time Step & $\Delta t$ & $\metre\second$ & $0.1$ & - & -	\\ \hline
				$\#$ of Sim. Steps & - & - & $100$ & - & -	\\ \hline
				Distance to RX & $d$ & $\mu\metre$ & Various & $0.01$ & $20$ \\ \hline
				TX Release Time & $t_0$	& $\metre\second$ & $0$ & $-10$ & $< t_1$  \\ \hline
				Diffusion Coefficient
				& $\Dx{}$ & $\metre^2/\second$ & $10^{-9}$ & $10^{-10}$ & $10^{-7}$ \\ \hline
				Degradation Rate & $\kth{}$ & $\second^{-1}$ & $62.5$ & $0$ &	$500$ \\ \hline
				Flow Towards RX & $\vpara$ & $\metre\metre/\second$ & $2$ & $-3$ & $6$ \\ \hline
				Perpendicular Flow & $\vperp$ & $\metre\metre/\second$ & $1$ & $0$ & $10$ \\ \hline
				Molecules Released & $\Nemit$ & - & $10^5$ & $10^3$ & $10^6$ \\ \hline
			\end{tabular}
		}
		\label{table_param}
	\end{table}
}

\newcommand{\figImpulse}[2]{
	\begin{figure}[#1]
		\centering
		\includegraphics[width=#2\linewidth]
		{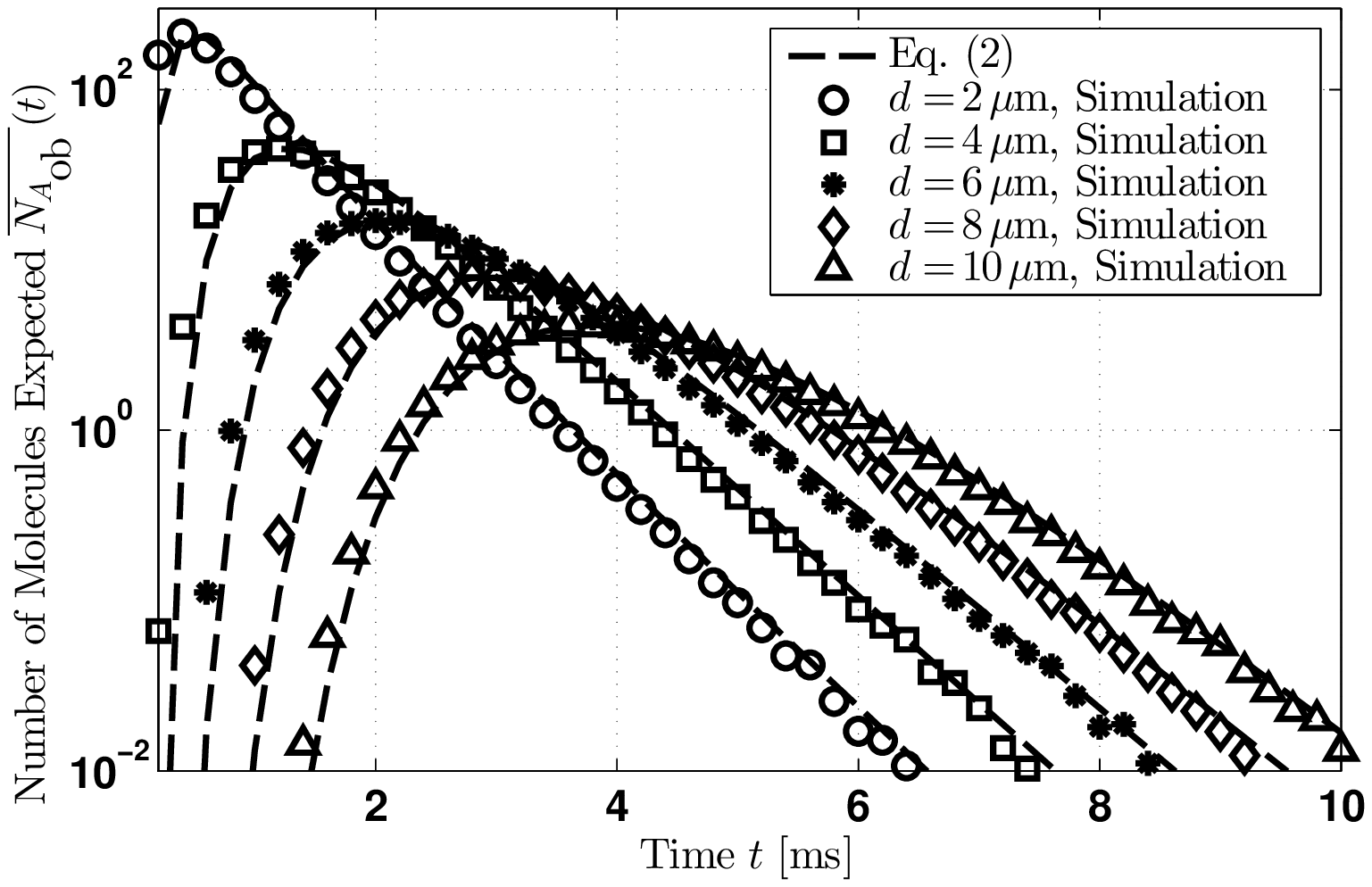}
		\caption{The expected channel impulse response $\Nxtavg{t}{ob}$ of the environment defined by Table~\ref{table_param} as a function of time $t$ for varying distance $d$. The responses in this figure are found by evaluating (\ref{EQ14_04_01}) and compared with corresponding simulations.}
		\label{fig_impulse}
	\end{figure}
}

\newcommand{\figUnknowns}[2]{
	\begin{figure}[#1]
		\centering
		\includegraphics[width=#2\linewidth]
		{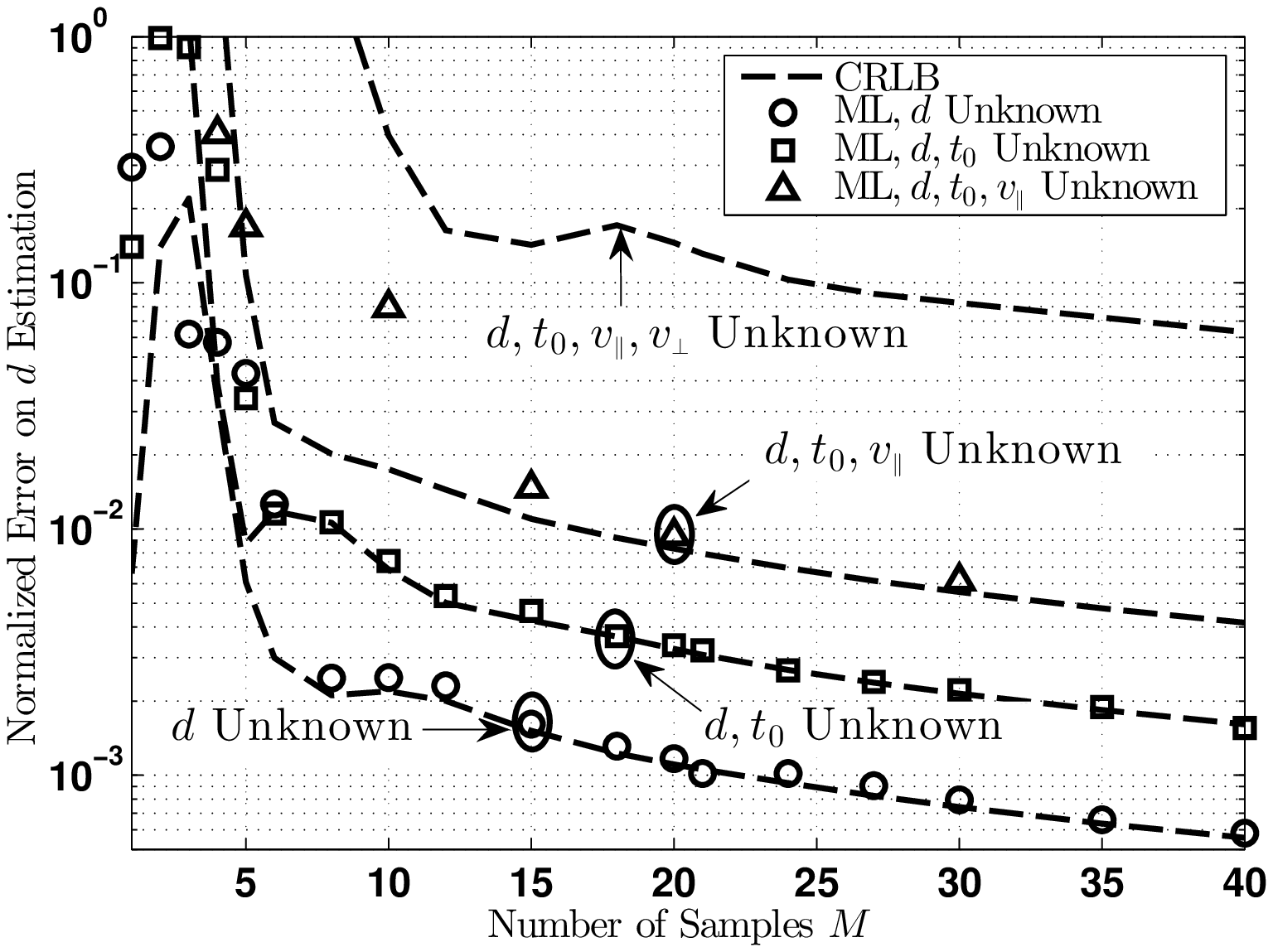}
		\caption{Normalized mean square error of ML distance estimation as a function of the number of observations $\M$ and as the knowledge of other parameters is removed. The corresponding CRLB for each estimate is also shown.}
		\label{fig_d_more_unknowns}
	\end{figure}
}

\newcommand{\figSingleAllML}[2]{
	\begin{figure}[#1]
		\centering
		\includegraphics[width=#2\linewidth]
		{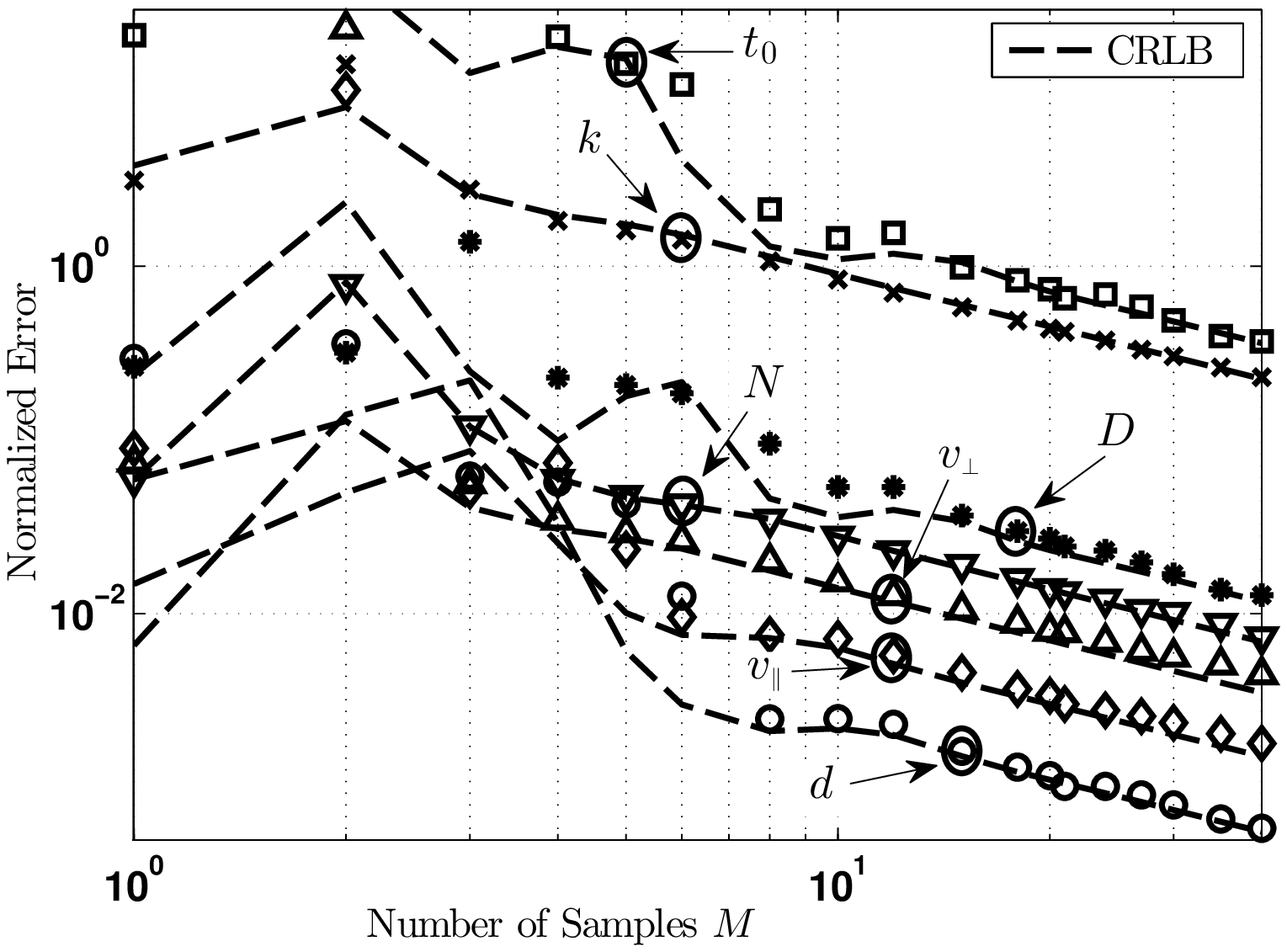}
		\caption{Normalized mean square error of ML estimation of each channel parameter when that parameter is the only one that is unknown. ML performance is given as a function of the number of observations $\M$ when the distance $d=6\,\mu\metre$. The corresponding CRLB for each estimate is also shown.}
		\label{fig_single_all}
	\end{figure}
}

\newcommand{\figAllDistUnknown}[2]{
	\begin{figure}[#1]
		\centering
		\includegraphics[width=#2\linewidth]
		{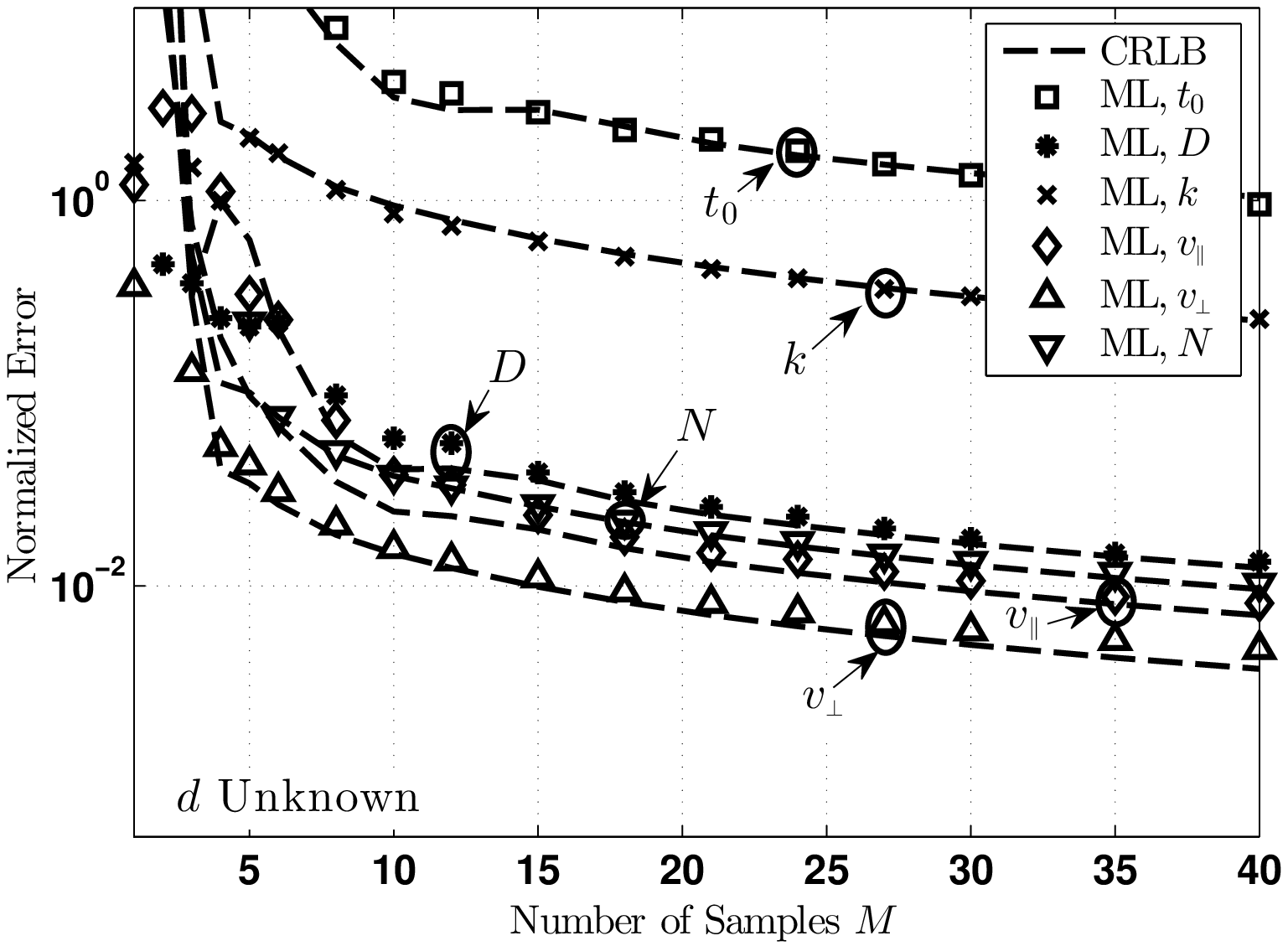}
		\caption{Normalized mean square error of ML estimation of each channel parameter when that parameter and the distance $d$ are unknown. ML performance is given as a function of the number of observations $\M$ when the distance $d=6\,\mu\metre$. The corresponding CRLB for each estimate is also shown.}
		\label{fig_all_d_unknown}
	\end{figure}
}

\newcommand{\figSingularity}[2]{
	\begin{figure}[#1]
		\centering
		\includegraphics[width=#2\linewidth]
		{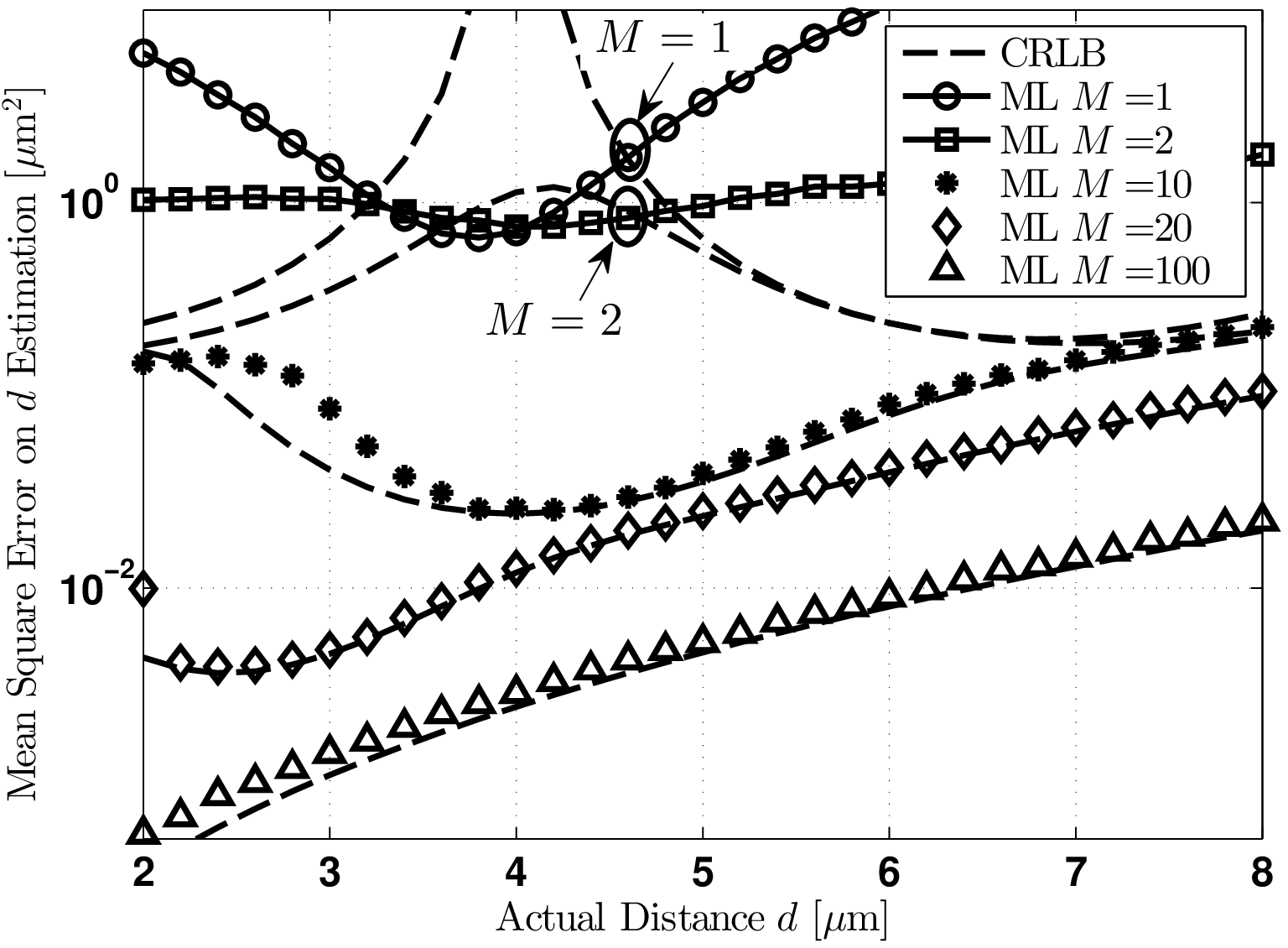}
		\caption{Mean square error of ML distance estimation when $d$ is the only unknown parameter. The corresponding CRLB for each estimate is also shown. For every value of $\M$, there is a sample taken at time $t_{\smM} =\,2\metre\second$.}
		\label{fig_d_singularity}
	\end{figure}
}

\newcommand{\figPeakEstAll}[2]{
	\begin{figure}[#1]
		\centering
		\includegraphics[width=#2\linewidth]
		{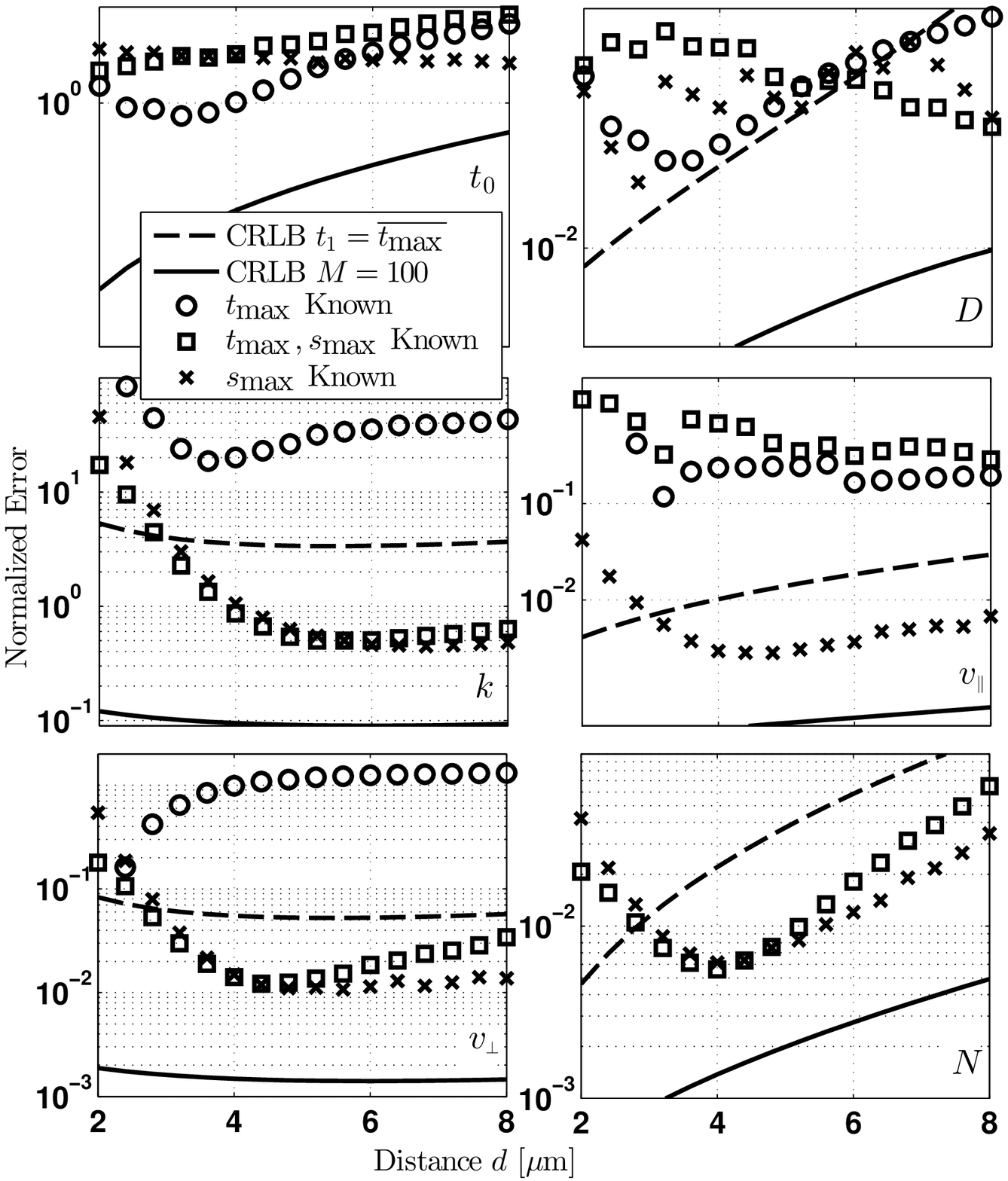}
		\caption{Normalized mean square error of peak-based estimation as a function of the distance $d$ when the window length is $\winLength=7$. Each subplot is labeled with the parameter being estimated. The CRLBs for $t_1 = \overline{\ttext{max}}$ and $\M=100$ are also shown. The release time $t_0$ does not have a CRLB when $t_1 = \overline{\ttext{max}}$ because there is \emph{always} a singularity at that time. The number of released molecules $\Nemit$ cannot be estimated from the knowledge of $\ttext{max}$ alone.}
		\label{fig_peak_all}
	\end{figure}
}

\newcommand{\figPeakEstDist}[2]{
	\begin{figure}[#1]
		\centering
		\includegraphics[width=#2\linewidth]
		{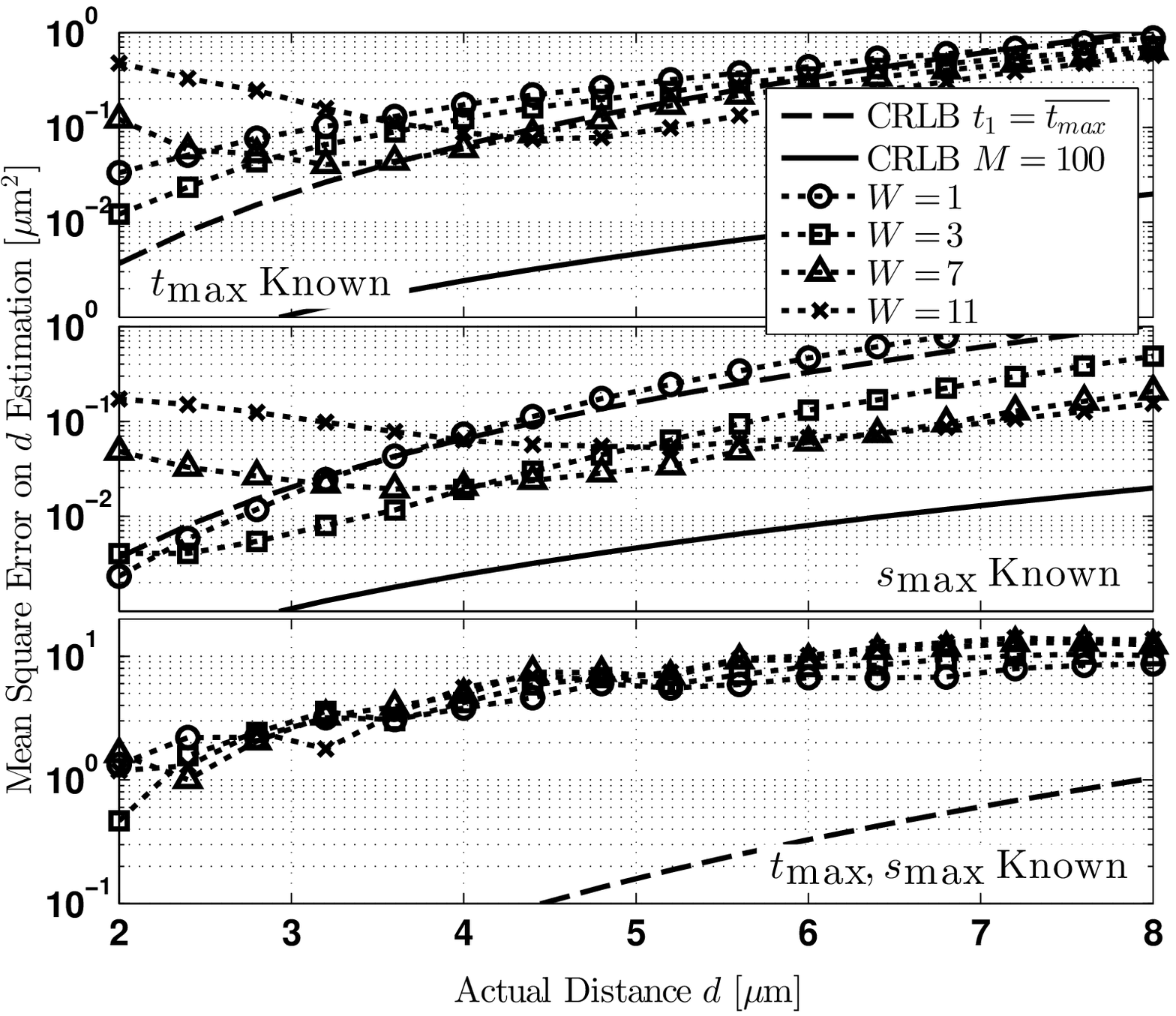}
		\caption{Mean square error of peak-based distance estimation as a function of the actual distance $d$ for varying window length $\winLength$. Each subplot is labeled with the knowledge available to the peak-based estimator. The CRLBs for $t_1 = \overline{\ttext{max}}$ and $\M=100$ are also shown and are the same in each subplot (although the CRLB when $\M=100$ in the bottom subplot is not visible on the scale shown).}
		\label{fig_peak_d}
	\end{figure}
}

\maketitle

\begin{abstract}
The design and analysis of diffusive molecular communication systems generally requires knowledge of the environment's physical and chemical properties. Furthermore, prospective applications might rely on the timely detection of changes in the local system parameters. This paper studies the local estimation of channel parameters for diffusive molecular communication when a transmitter releases molecules that are observed by a receiver. The Fisher information matrix of the joint parameter estimation problem is derived so that the Cramer-Rao lower bound on the variance of locally unbiased estimation can be found. The joint estimation problem can be reduced to the estimation of any subset of the channel parameters. Maximum likelihood estimation leads to closed-form solutions for some single-parameter estimation problems and can otherwise be determined numerically. Peak-based estimators are proposed for low-complexity estimation of a single unknown parameter.
\end{abstract}

\begin{IEEEkeywords}
Cramer-Rao lower bound, diffusion, maximum likelihood estimation, molecular communication, parameter estimation.
\end{IEEEkeywords}

\section{Introduction}

\PARstart{M}{olecular} communication (MC) is the transmission of information where molecules are used as information carriers. MC is ubiquitous in biological systems. For example, endocrine signaling is the release of hormone molecules that propagate via the bloodstream, paracrine signaling is the release of molecules into extracellular fluid that are detected by local cells, and molecules are also released in the synapses between neurons to relay signals between them; for more details, see \cite[Ch. 16]{RefWorks:588}. Despite their widespread use, MC systems in nature are typically designed for the transmission of limited quantities of information, e.g., a message that is a time-varying ON/OFF control signal for a biological process.
Recent advances in nanotechnology have motivated interest in synthetic communication networks where the principles of MC are used to deliver arbitrary amounts of information in environments where the deployment of traditional wireless communication networks is unsafe or infeasible. These networks could advance applications in a diverse number of fields, including biological engineering, healthcare, and manufacturing; see \cite{RefWorks:801}.

Diffusion-based MC relies on the random motion of information molecules due to collisions with other molecules in the propagation environment. When molecules are released into a diffusive environment, they can be transported from a transmitter to its corresponding receiver without any additional infrastructure or external energy. However, this is an imperfect process that can be best described by an \emph{expected} channel impulse response, i.e., the number of molecules \emph{expected} at a receiver when molecules are released at some instant by a transmitter. The expected channel impulse response is a function of the parameters of the diffusive environment, including its geometry, the distance from the transmitter to the receiver, the diffusion coefficient of the molecules, and the time elapsed since the molecules were released. Other phenomena can also impact the status of the diffusing molecules and hence the channel impulse response. These phenomena include chemical reactions that have the molecules of interest as a product or reactant, other sources of those molecules that are not the intended transmitter, and whether there is any bulk fluid flow.

Given that the channel impulse response depends on the environmental parameters, the response can be used as a local noisy observation to estimate the values of those parameters. This is especially true when the expected impulse response can be written in closed form (although simplifying assumptions are generally needed to obtain a closed-form expression). By observing the arrival of molecules from a transmitter, an intelligent receiver might learn about the current local conditions of the propagation environment, which is essential for some prospective MC applications.

For example, consider a healthcare application where a network of microscale sensors are deployed to monitor a patient's bloodstream. The sensors might need to be mounted at regular intervals along the blood vessel walls, such that they need to estimate the distance separating themselves before mounting. By monitoring the remaining individual channel parameters, changes could be detected and the cause of the change might be inferred. The blood flow velocity could be a proxy for blood pressure. The diffusion coefficient could be a proxy for blood composition and used to identify major changes in blood cell counts, as described in \cite{RefWorks:868}. The chemical kinetics of the information molecules could be a proxy for blood pH; chemical reactivity varies with pH, as discussed in \cite[Ch.~10]{RefWorks:585}. In summary, knowledge of the individual channel parameters can be more insightful than knowledge of the expected channel impulse response alone, with the caveat that estimating individual parameters is only feasible if an expression for the channel impulse response as a function of the parameters is available.

We note that there are also macroscale estimation methods that are used to measure channel parameters. For example, there are various experimental methods to measure fluid diffusion coefficients, such as diaphragm cells and Taylor dispersion; see \cite[Ch.~5]{RefWorks:742}. However, these methods are appropriate for laboratory environments and might not be suitable for on-going measurements in confined settings where the deployment of an MC system might be less invasive.

In this paper, we study local joint channel parameter estimation in a diffusive MC environment, where in the most general case we assume that we know the form of the expression for the expected channel impulse response but we assume that we know \emph{none} of the individual parameter values. Specifically, we consider the system model that we studied in \cite{RefWorks:747}, where a fixed receiver in an unbounded 3-dimensional environment observes molecules released by a fixed impulsive source. The molecules experience steady uniform flow and can probabilistically degrade. We ignore the presence of other molecule sources. For tractability, the receiver is a passive observer that can perfectly count the number of information molecules within its volume at a given instant. Each count is an observation and one or multiple observations are used to estimate the parameters.

We study an ideal model for two reasons. First, to the best of our knowledge it is the most detailed diffusive MC model for which a closed-form time domain expression of the channel impulse response is available. Second, the characteristics of this model approximate special cases of more realistic environments. For example, an environment that is sufficiently large relative to the distance between communicating devices can be assumed to be infinitely large. Also, an impulsive point source is sufficient to approximate a larger source that releases molecules sufficiently fast relative to the time required for molecules to reach the receiver via diffusion and flow (in fact, we use a non-point source in our simulations with negligible impact on estimation performance). In summary, estimator performance within this ideal model can serve as a bound or benchmark for performance in more realistic environments.

Existing literature on parameter estimation via diffusive MC has been limited to one unknown parameter. The distance between devices has been estimated in \cite{RefWorks:808, RefWorks:802,RefWorks:614,RefWorks:776,RefWorks:674}, whereas the time of transmitter release (i.e., synchronization) has been estimated in \cite{RefWorks:761,RefWorks:773}. With the exception of our preliminary work in \cite{RefWorks:808}, parameter estimation has only been considered in environments with diffusion alone and not with fluid flow or molecule degradation.

In our model, when the transmitter releases an impulse of molecules, the unknown parameters are the time that the molecules are released, the number of molecules released, the distance to the receiver, the diffusion coefficient, the fluid flow vector, and the molecule degradation rate. We are interested in determining the best possible performance of the (classical\footnote{We focus here on classical approaches, where we assume no prior knowledge about the probability distribution of the parameters being estimated. Bayesian approaches assume that the unknown parameter is sampled from a known distribution; see \cite[Ch.~10]{RefWorks:803}. We leave the study of such approaches for future work.}) joint estimation of \emph{all} of these parameters, as a function of the observations made by the receiver. We aim to provide bounds on the performance of any estimation protocol. We do not claim that estimating all parameters simultaneously is practical. Rather, our analysis easily simplifies to the estimation of \emph{any subset} of the channel parameters. For example, our analysis of distance estimation in \cite{RefWorks:808} is a special case of the complete analysis that we present here. Furthermore, we gain insight into how the knowledge of any one parameter decreases the error in estimating any of the other parameters. The primary contributions of this paper are summarized as follows:
\begin{enumerate}
	\item We derive the Fisher Information Matrix (FIM) of our joint parameter estimation problem to give the Cramer-Rao lower bound (CRLB) on the variance of estimation error of any locally \emph{unbiased} estimator as a function of independent observations of a transmitted impulse. Bounds on the unbiased estimation of any subset of the channel parameters can be found by considering only the corresponding elements of the FIM (e.g., if only estimating the distance, as we did in \cite{RefWorks:808}, then only 1 of the 28 unique terms in the FIM is needed).
	\item We study maximum likelihood (ML) estimation of our joint parameter estimation problem. Closed-form solutions exist for some single-parameter estimation problems with one observation, as we showed for distance estimation in \cite{RefWorks:808}. Otherwise, ML estimates can be determined numerically, via either the Newton-Raphson method or an exhaustive search.
	\item We consider the presence of singularities in the FIM, in which case any unbiased estimator will have infinite variance. Dealing with singularities is an open problem in the parameter estimation literature, cf. e.g. \cite{RefWorks:866,RefWorks:853,RefWorks:854,RefWorks:855}. Singularities in the FIM, or being in the ``vicinity'' of a singularity, can have an impact when estimating one parameter or multiple parameters simultaneously.
	\item We propose peak-based estimators for low-complexity estimation of a single parameter. Variants of peak-based distance estimators were originally presented in \cite{RefWorks:614,RefWorks:776}. We present a comprehensive discussion of how the peak molecule observation and/or the time of the peak number of observed molecules can be used to estimate any single parameter, given knowledge of the other parameters.
\end{enumerate}

We note that we focus on parameter estimation when there is only \emph{one} device releasing molecules, i.e., the transmitter, and they are observed by the receiver. We coined the term \emph{one-way} protocols in \cite{RefWorks:808} to refer to estimation protocols using this approach, and to distinguish them from \emph{two-way} protocols (such as those proposed for distance estimation in \cite{RefWorks:802,RefWorks:614,RefWorks:674}), which rely on feedback from the receiver back to the transmitter so that the transmitter makes the estimate. In general, two-way protocols can be no more accurate than one-way protocols, because two-way methods require the subsequent detection of two molecule impulses.

The rest of this paper is organized as follows. In Section~\ref{sec_model}, we describe the physical environment, review the expected channel impulse response, and review the CRLB and ML estimation. We derive the FIM of the joint estimation problem, from which the CRLB can be found, in Section~\ref{sec_joint_est}. In Section~\ref{sec_protocol}, we apply examples of ML estimation to the joint estimation problem and present the peak-based estimation protocols. We present numerical and simulation results in Section~\ref{sec_num}. Conclusions are drawn
in Section~\ref{sec_concl}.

\section{System Model and Estimation Preliminaries}
\label{sec_model}

In this section, we describe the diffusive environment and the expected channel impulse response. We review the definition of the CRLB for vector parameter estimation. We also review ML estimation and the Newton-Raphson method for numerical evaluation of the ML estimate.

\subsection{Physical Environment}

We consider a 3-dimensional fluid environment as shown in Fig.~\ref{fig_model}. The environment is unbounded and with uniform temperature and viscosity. There are two fixed devices, which we label the transmitter (TX) and the receiver (RX) because we focus on \emph{one-way} parameter estimation. The TX is a point that is distance $d$ from the center of the RX. The RX is a sphere of radius $\rrx$ and volume $\Vrx$. The coordinate axes are defined by placing the center of the RX at the origin and the TX at Cartesian coordinates $\{-d,0,0\}$. As we noted in \cite{RefWorks:747}, concentrations observed in this environment are equivalent by a factor of two to those in the semi-infinite case where the $\x\y$-plane is an elastic boundary and the RX is a hemisphere; see \cite[Eq.~(2.7)]{RefWorks:586}. There is a steady uniform flow $\vxvec{}$ with components $\vpara$ and $\vperp$. $\vpara$ is the component of $\vxvec{}$ in the direction of a line pointing from the TX towards the RX, and $\vperp$ is the component of $\vxvec{}$ perpendicular to $\vpara$ (the precise direction of $\vperp$ is irrelevant due to symmetry). We note that uniform flows are the simplest analytically but do not generally describe the flow in cylindrical environments such as blood vessels, where flows are described as laminar (where successive layers of fluid slide over one another without mixing) or turbulent (where fluid motion is even more chaotic than under diffusion alone), depending on the relative importance of inertial and viscous forces; see \cite[Ch.~2]{RefWorks:588}.

\ifOneCol

	\begin{figure}[!tb]
		\centering
		\def\svgwidth{0.55\linewidth}
		\input{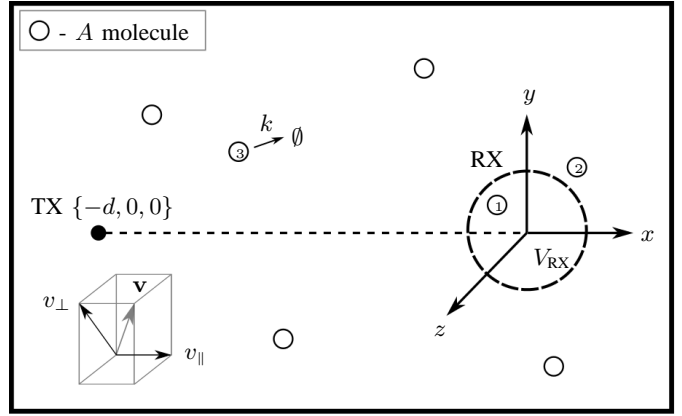}
		\caption{The system model considered throughout this paper. The TX is a point source of $\A$ molecules and the RX is a passive observer centered at the origin. The $\A$ molecules are shown as small hallow circles and some are labeled. Molecule $1$ is inside $\Vrx$ and so can be observed by the RX. Molecule $2$ was previously inside $\Vrx$ and is now outside because the RX is non-absorbing. Once released by the TX, the behavior of each molecule is that of a biased random walk (biased by the steady flow $\vxvec{}$) until it undergoes degradation via the chemical reaction described by first-order degradation rate constant $\kth{}$, e.g., molecule $3$.}
		\label{fig_model}
	\end{figure}

\else

	\begin{figure}[!tb]
		\centering
		\def\svgwidth{1\linewidth}
		\input{graphics/2014_09_system_model.eps_tex}
		\caption{The system model considered throughout this paper. The TX is a point source of $\A$ molecules and the RX is a passive observer centered at the origin. The $\A$ molecules are shown as small hallow circles and some are labeled. Molecule $1$ is inside $\Vrx$ and so can be observed by the RX. Molecule $2$ was previously inside $\Vrx$ and is now outside because the RX is non-absorbing. Once released by the TX, the behavior of each molecule is that of a biased random walk (biased by the steady flow $\vxvec{}$) until it undergoes degradation via the chemical reaction described by first-order degradation rate constant $\kth{}$, e.g., molecule $3$.}
		\label{fig_model}
	\end{figure}

\fi

The TX is a source of molecules, labeled $\A$ molecules, that can be detected by the RX. The $\A$ molecules independently diffuse with constant diffusion coefficient $\Dx{}$, and they can degrade anywhere in the propagation environment via a first-order chemical reaction that can be written as
\begin{align}
\label{k_mechanism}
& \A \xrightarrow{\kth{}} \emptyset,
\end{align}
where $\kth{}$ is the first-order reaction rate constant in $\second^{-1}$. We do not specify the product of reaction (\ref{k_mechanism}), except to say that it is not recognizable by the RX. We ignore the reaction kinetics of the reception process at the RX for tractability (recently, the time domain channel impulse response for a similar but simpler system model was derived in \cite{RefWorks:844}). Instead, the RX is a passive observer that can perfectly count the number of $\A$ molecules within its volume $\Vrx$ at any desired time, i.e., the molecules are observed without being bound or consumed.

Given our system model, we can write the expected channel impulse response. We assume that the concentration of $\A$ molecules expected at the RX due to a release of $\A$ molecules by the TX is uniform throughout the RX and equal to that expected at the center of the RX. We previously studied the accuracy of this assumption in environments with molecule degradation in \cite{RefWorks:706} and in flowing environments in \cite{RefWorks:752}. In both works, we showed that this assumption is accurate for an RX that is sufficiently small relative to its distance from the TX. Using this assumption, if the TX instantaneously releases $\Nemit$ $\A$ molecules at time $t = t_0$, then the number of those molecules expected to be observed by the RX at time $t$, $\Nxtavg{t}{ob}$, is given by
\cite[Eq. (12)]{RefWorks:747}
\ifOneCol
\begin{equation}
\label{EQ14_04_01}
\Nxtavg{t}{ob} = \frac{\Nemit\Vrx}{(4\pi \Dx{}
	(t - t_0))^{3/2}}\EXP{-\kth{}(t - t_0) - \frac{\radmag{ef}^2}{4\Dx{} (t - t_0)}},
\end{equation}
\else
\begin{align}
\label{EQ14_04_01}
\Nxtavg{t}{ob} = &\; \frac{\Nemit\Vrx}{(4\pi \Dx{}
	(t - t_0))^{3/2}} \nonumber \\
	& \times\EXP{-\kth{}(t - t_0) - \frac{\radmag{ef}^2}{4\Dx{} (t - t_0)}},
\end{align}
\fi
where $\radmag{ef}^2 = (d - \vpara (t - t_0))^2 + (\vperp (t - t_0))^2$ is the
square of the \emph{effective} distance from the TX to the RX. For compactness, we define $\ttext{ef} = t - t_0$ as the elapsed time since the molecules were released, i.e., $\ttext{ef}>0$. We note that (\ref{EQ14_04_01}) can be derived as an extension of \cite[Eq.~(107)]{RefWorks:940} by modifying the underlying differential equation to account for flow and molecule degradation, as we did in \cite{RefWorks:747} and \cite{RefWorks:662}, respectively. The actual number of molecules observed by the RX is $\Nobst{t}$, and the time-varying mean of $\Nobst{t}$ is given by (\ref{EQ14_04_01}). The only variable in (\ref{EQ14_04_01}) that we always assume is known to the RX is its volume $\Vrx$. We summarize the remaining channel parameters in Table~\ref{table_fim}, and we assume that some subset of those parameters are unknown and must be estimated.

\ifOneCol
\tableFIMOneCol{!tb}
\else
\tableFIMTwoCol{!tb}
\fi

\subsection{The Cramer-Rao Lower Bound}
\label{sec_CRLB_def}

The Cramer-Rao lower bound is a bound on the error variance of any (locally) \emph{unbiased} estimator; biased estimators, or estimators that are locally biased, can outperform the CRLB. Here, we review the definition of the CRLB for a vector parameter as described in \cite[Ch. 3]{RefWorks:803}. The definition easily simplifies in the case of a single unknown parameter.

Assume that we have a vector of $\M$ observations $\vect{\sx{}} = [\sx{1},\ldots,\sx{\M}]^T$ and a vector of $\numParam$ unknown parameters $\vectm{\theta} = [\theta_1, \ldots,\theta_\numParam]^T$, where $[\cdot]^T$ is vector transpose. Assume that we know the conditional probability density function (PDF) of the observations, $p(\vect{\sx{}}|\vectm{\theta})$.
Under standard regularity conditions (see \cite[Ch.~1.7]{RefWorks:942}), and by \cite[Th. 3.2]{RefWorks:803}, the covariance matrix of any unbiased estimator for $\vectm{\theta}$, $\mathbf{C}_{\hat{\vectm{\theta}}}$, satisfies
\begin{equation}
\label{EQ14_04_95}
\mathbf{C}_{\hat{\vectm{\theta}}} - \mathbf{I}^{-1}\left(\vectm{\theta}\right) \ge \vect{0},
\end{equation}
where $\ge \vect{0}$ means that the matrix is positive semi-definite. An estimator is unbiased if $E[\hat{\vectm{\theta}}] = \vectm{\theta}$. The elements of the Fisher information matrix $\mathbf{I}\left(\vectm{\theta}\right)$ are given by
\begin{equation}
\label{EQ14_04_96}
\left[\mathbf{I}\left(\vectm{\theta}\right)\right]_{\theta_i,\theta_j} =
-E\left[\psbypxpy{\ln p(\vect{\sx{}}|\vectm{\theta})}{\theta_i}{\theta_j}\right],
\end{equation}
where $E\left[\cdot\right]$ is the expectation taken with respect to
$p(\vect{\sx{}}|\vectm{\theta})$, and the derivatives are evaluated at the true value of $\vectm{\theta}$. For a positive semi-definite matrix, the diagonal elements are non-negative. Thus, from (\ref{EQ14_04_95}) we have
\begin{equation}
\label{EQ14_04_97}
\left[\mathbf{C}_{\hat{\vectm{\theta}}} - \mathbf{I}^{-1}\left(\vectm{\theta}\right)\right]_{\theta_i,\theta_i} \ge 0,
\end{equation}
and
\begin{equation}
\label{EQ14_04_98}
\var{\hat{\theta}_i} = 
\left[\mathbf{C}_{\hat{\vectm{\theta}}}\right]_{\theta_i,\theta_i} \ge \left[\mathbf{I}^{-1}\left(\vectm{\theta}\right)\right]_{\theta_i,\theta_i},
\end{equation}
where $\var{\hat{\theta}_i}$ is defined as the variance of the estimation error of parameter $\theta_i$, i.e.,
\begin{equation}
\label{var_def}
\var{\hat{\theta}_i} = E\!\left[\!(\hat{\theta}_i - E[\hat{\theta}_i])^2\!\right].
\end{equation}

Thus, the CRLB on the error variance of the $i$th parameter, when all $\numParam$ parameters are jointly estimated by an unbiased estimator, is given by the $i$th diagonal element of the inverse of the FIM. The elements of the FIM are found using (\ref{EQ14_04_96}).

\subsection{Maximum Likelihood Estimation}
\label{sec_ml_def}

ML estimation is known as a ``turn-the-crank'' procedure because it can be procedurally implemented for many estimation problems where the observation PDF is known; see \cite[Ch.~7]{RefWorks:803} and examples of exceptions in \cite[Ch.~6]{RefWorks:941}. It is generally accepted that, in most cases, ML estimation is asymptotically efficient in the sense of the CRLB as the number of observations grows large, i.e., as $\M \to \infty$; see \cite[Ch.~6]{RefWorks:941}. However, we cannot make any general claims about the bias or the relative performance of ML estimation for a \emph{finite} number of observations.

The ML estimate of vector parameter $\vectm{\theta}$ is given as follows:
\begin{equation}
\label{mle_def}
\hat{\vectm{\theta}} \textBar{ML} = \argmax_{\vectm{\theta}} p(\vect{\sx{}}|\vectm{\theta}),
\end{equation}
i.e., the ML estimate of $\vectm{\theta}$ is the vector that maximizes the observation PDF, given the observation vector $\vect{\sx{}}$. We will find that there are special cases, particularly if there is one observation and one unknown parameter, where we can write the ML estimate in closed form. In general, it can be found numerically. For example, we can consider the Newton-Raphson method to avoid performing an exhaustive search (the latter becomes computationally cumbersome when there are multiple unknown parameters). The Newton-Raphson method begins with an initial estimate $\hat{\vectm{\theta}}_0$. The $(n+1)$th estimate is found iteratively as \cite[Eq. (7.48)]{RefWorks:803}
\begin{equation}
\label{newton_raphson_def}
\hat{\vectm{\theta}}_{n+1} = \hat{\vectm{\theta}}_n - \left[\psbypxpy{\ln p(\vect{\sx{}}|\vectm{\theta})}{\vectm{\theta}}{\vectm{\theta}^T}\right]^{-1} \pbypx{\ln p(\vect{\sx{}}|\vectm{\theta})}{\vectm{\theta}} \bigg|_{\vectm{\theta} = \hat{\vectm{\theta}}_n},
\end{equation}
where
\begin{equation}
\label{newton_raphson_def_2}
\left[\psbypxpy{\ln p(\vect{\sx{}}|\vectm{\theta})}{\vectm{\theta}}{\vectm{\theta}^T}\right]_{i,j} = \psbypxpy{\ln p(\vect{\sx{}}|\vectm{\theta})}{\theta_i}{\theta_j} \quad \forall i,j \in \{1,\ldots, \numParam\}.
\end{equation}

The convenience in implementing the Newton-Raphson method is that the expressions for the derivatives required in (\ref{newton_raphson_def}) and (\ref{newton_raphson_def_2}) can be found while deriving the elements of the FIM in (\ref{EQ14_04_96}). We will provide examples of this procedure in Section~\ref{sec_ml}. We must also recognize the limitations of the Newton-Raphson method, as discussed in \cite[Ch. 7]{RefWorks:803}. The method is not guaranteed to converge, or it might converge to a \emph{local} maximum. The method can quickly diverge if the current estimate results in an FIM that is close to singular. Generally, the ML estimate will be found if the initial estimate $\hat{\vectm{\theta}}_0$ is close to the ML estimate and not in the ``vicinity'' of singularities (we discuss the meaning of being in the vicinity of a singularity in the FIM in further detail in Section~\ref{sec_ml}).

\section{Joint Parameter Estimation Performance}
\label{sec_joint_est}

In this section, we first derive the FIM of the joint parameter estimation problem in diffusive MC with steady uniform flow and first-order molecule degradation. Then, we present simple examples of how to use the FIM to find the CRLB (following the methodology in Section~\ref{sec_CRLB_def}) and comment on situations where the FIM is singular, i.e., where the CRLB does not exist.

\subsection{Main Result}
\label{sec_fim}
To derive the FIM, we first need the joint observation PDF $p(\vect{\sx{}},\vectm{\theta})$ for our problem. The TX makes a single release of $\Nemit$ molecules at time $t = t_0$. Our observations are the discrete number of molecules found within $\Vrx$ at the sampling times, i.e., $\sx{\smM} = \Nobst{t_\smM}$, where the $\smM$th observation is made at time $t_\smM$. We assume that the time between successive observations is sufficient for each observation $\sx{\smM}$ to be independent (we discussed the independence of observations in detail in \cite{RefWorks:747}). We will also assume that the individual observations, which are Binomially distributed, can be approximated as Poisson random variables whose means are the expected values of the observations at the corresponding times (this has been shown to be highly accurate in our previous work, including \cite{RefWorks:747,RefWorks:662}, although the Gaussian approximation can become more accurate as it becomes more likely to observe any individual molecule). Thus, the joint PDF is
\cite[Eq.~(9)]{RefWorks:808}
\begin{equation}
\label{EQ14_04_07}
p(\vect{\sx{}}|\vectm{\theta}) = \prod_{\smM = 1}^\M
{\Nxtavg{t_{\smM}}{ob}}^{\sx{\smM}}\EXP{-\Nxtavg{t_{\smM}}{ob}}/\sx{\smM}!,
\end{equation}
where $\Nxtavg{t_{\smM}}{ob}$ is as given by (\ref{EQ14_04_01}). The logarithm of the joint PDF is
\begin{equation}
\label{EQ14_04_93}
\ln p(\vect{\sx{}}|\vectm{\theta}) = \sum_{\smM = 1}^\M \left[
\sx{\smM}\ln\Nxtavg{t_{\smM}}{ob}-\ln\sx{\smM}! - \Nxtavg{t_{\smM}}{ob}\right].
\end{equation}

We summarize the channel parameters that we wish to estimate in Table~\ref{table_fim}. From (\ref{EQ14_04_93}) and (\ref{EQ14_04_01}), the FIM can be found. We present the final result in the following theorem:
\begin{theorem}[FIM of the Joint Estimation Problem]
	\label{theorem_fim}
	The elements of the Fisher information matrix for the joint parameter estimation problem are of the form
	\begin{equation}
	\label{EQ14_04_107_gen}
	\left[\mathbf{I}\left(\vectm{\theta}\right)\right]_{\theta_i,\theta_j} = \sum_{\smM = 1}^\M G_{\theta_i}G_{\theta_j}\Nxtavg{t_{\smM}}{ob},
	\end{equation}
	where $G_{\theta_i}$ is a unique term for parameter $\theta_i$ and we note that the ordering of the elements in $\mathbf{I}\left(\vectm{\theta}\right)$ is arbitrary. The $G_{\theta_i}$ terms for the channel parameters are as follows:
	\begin{align}
	\label{EQ14_04_156}
	G_{d} = &\; 
	\frac{1}{2\Dx{}}\left(\vpara - \frac{d}{\ttext{ef}} \right), \\
	\label{EQ14_04_157}
	G_{t_0} = &\; 
	\left(\frac{3}{2\ttext{ef}} + \kth{} + \frac{\vpara^2 + \vperp^2}{4\Dx{}} - \frac{d^2}{4\Dx{}\ttext{ef}^2} \right), \\
	\label{EQ14_04_158}
	G_{\Dx{}} = &\; \frac{1}{2\Dx{}}
	\left[\frac{1}{2\Dx{}}\left(\frac{d^2}{\ttext{ef}} - 2d\vpara +\ttext{ef}\left(\vpara^2 + \vperp^2 \right) \right) - 3 \right], \\
	\label{EQ14_04_159}
	G_{\kth{}} = &\;
	-\ttext{ef}, \\
	\label{EQ14_04_160}
	G_{\vpara} = &\; \frac{1}{2\Dx{}}
	\left(d - \vpara \ttext{ef} \right), \\
	\label{EQ14_04_161}
	G_{\vperp} = &\; -\frac{\vperp \ttext{ef}}{2\Dx{}}
	, \\
	\label{EQ14_04_162}
	G_{\Nemit} = &\; \frac{1}{\Nemit},
	\end{align}
	where here $\ttext{ef} = t_{\smM} - t_0$. The diagonal elements of the FIM are presented in Table~\ref{table_fim}, such that $\left[\mathbf{I}\left(\vectm{\theta}\right)\right]_{\theta_i}$ is the diagonal element associated with parameter $\theta_i$. The 21 unique off-diagonal elements can be analogously found from (\ref{EQ14_04_107_gen}).
\end{theorem}
\begin{IEEEproof}
	The proof is straightforward by applying the properties of logarithms and exponentials and the rules of differentiation\footnote{An alternative (equivalent) derivation can be made directly from (\ref{EQ14_04_07}) and (\ref{EQ14_04_01}), as identified by an anonymous reviewer. We can recognize that the Fisher information of the mean of a single Poisson distribution is the inverse of that mean, and then apply the commutative property of Fisher information for independent Poisson distributions and the reparametrization rule for Fisher information; see \cite[Ch.~2]{RefWorks:941}.} to (\ref{EQ14_04_93}) and (\ref{EQ14_04_01}), and by noting that (by definition) $E\left[\sx{\smM}\right] = \Nxtavg{t_{\smM}}{ob}$. It can be shown that the $G_{\theta_i}$ terms come from the derivative of the logarithm of the joint PDF in (\ref{EQ14_04_93}) with respect to $\theta_i$, i.e.,
	\begin{equation}
	\label{EQ14_04_156_gen}
	\pbypx{\ln p(\vect{\sx{}}|\vectm{\theta})}{\theta_i} = 
	\sum_{\smM = 1}^\M G_{\theta_i}\left(\sx{\smM} - \Nxtavg{t_{\smM}}{ob} \right).
	\end{equation}

\end{IEEEproof}

\subsection{Examples of the CRLB}
\label{sec_crlb_ex}

The size of the FIM for a specific problem depends on the number of unknown channel parameters, i.e., given that there are $\numParam$ unknown parameters, the FIM will be an $\numParam \times \numParam$ matrix. The size of the FIM does \emph{not} depend on the number of parameters that we \emph{want} to estimate. If we want to estimate $\numEst$ parameters, then we should have $\numEst \le \numParam$. Here, we present two basic examples of using the FIM to find the CRLB. We consider estimating the distance $d$, then jointly estimating $d$ and the molecule release time $t_0$, because the distance between any pair of devices in the same environment can be unique, and every device can have its own internal synchronization. Thus, $d$ and $t_0$ are arguably the most critical parameters when establishing a communication link between a pair of devices.

The simplest scenario is the estimation of a single parameter when we assume that all other parameters are known. We studied this case for distance estimation in \cite{RefWorks:808}. By (\ref{EQ14_04_98}), we see that we only need to invert the corresponding entry in Table~\ref{table_fim}, and we can write the lower bound on the variance of any unbiased distance estimator as \cite[Th.~1]{RefWorks:808}
\begin{equation}
\label{EQ14_04_10}
\var{\hat{d}} \ge \frac{4{\Dx{}}^2}
{\sum_{\smM =
		1}^\M\left(\vpara-\frac{d}{\ttext{ef}}\right)^2\Nxtavg{t_{\smM}}{ob}}.
\end{equation}

Similarly, the FIM for any one unknown parameter has a single element and it can be easily inverted to find the CRLB. As we discussed in \cite{RefWorks:808}, equations for the CRLB give us insight into the factors that affect the accuracy of an estimate. For example, from (\ref{EQ14_04_10}) we see that a more accurate estimate might be possible if more samples are taken, i.e., by increasing $\M$. The same observation can be made for the estimation of any single parameter via inspection of the diagonal elements of the FIM in Table~\ref{table_fim}. The impact of some parameters, such as $\Dx{}$ on the estimation of $d$, or $t_{\smM}$ on the estimation of the degradation rate, are not immediately clear because the parameters are both inside and outside the $\Nxtavg{t_{\smM}}{ob}$ term in the corresponding FIM element. However, we can see from Table~\ref{table_fim} that increasing the number of molecules $\Nemit$ will also increase the bound on the variance of estimation of $\Nemit$, because $\Nxtavg{t_{\smM}}{ob}$ is only scaled by a factor of $\Nemit$.

For $\numParam > 1$, we must perform a matrix inversion to obtain the CRLB. Consider $\numParam = 2$ where $\vectm{\theta} = [d, t_0]^T$. The structure of the FIM is then
\begin{equation}
\label{FIM_ex2}
\mathbf{I}\left(\vectm{\theta}\right) = \begin{bmatrix*}[l]
\left[\mathbf{I}\left(\vectm{\theta}\right)\right]_{d} & \left[\mathbf{I}\left(\vectm{\theta}\right)\right]_{d,t_0} \\
\left[\mathbf{I}\left(\vectm{\theta}\right)\right]_{d,t_0} & \left[\mathbf{I}\left(\vectm{\theta}\right)\right]_{t_0}
\end{bmatrix*},
\end{equation}
where $\left[\mathbf{I}\left(\vectm{\theta}\right)\right]_{d}$, $\left[\mathbf{I}\left(\vectm{\theta}\right)\right]_{t_0}$ are from Table~\ref{table_fim}, and $\left[\mathbf{I}\left(\vectm{\theta}\right)\right]_{d,t_0}$ can be evaluated from (\ref{EQ14_04_107_gen}) using (\ref{EQ14_04_156}) and (\ref{EQ14_04_157}). The inversion of (\ref{FIM_ex2}) is straightforward. For brevity, we omit writing the inversion out in full, but we have two comments regarding its use. First, we did not need to specify \emph{which} parameter(s) we are trying to estimate, i.e., the FIM in (\ref{FIM_ex2}) applies to estimating $d$ or $t_0$ or both, given that both are unknown. Second, it can be shown that the CRLB for either parameter cannot be smaller than if that parameter were the \emph{only} unknown parameter. These two comments apply to any joint estimation problem (see \cite[Ch.~3]{RefWorks:803}); the FIM depends on the $\numParam$ unknown parameters and not the parameters being actively estimated, and the CRLB never decreases when more parameters become unknown. We show more examples of these observations when we present our numerical results in Section~\ref{sec_num}.

\subsection{On the Nonexistence of the CRLB}
\label{sec_singular}

Our analysis and discussion of the CRLB would be incomplete if we did not consider the occasions when the CRLB does not exist. By inspection of (\ref{EQ14_04_107_gen}) when there is a single observation, i.e., $\M=1$, we can see that singularities arise when $G_{\theta_i} = 0$, such that inversion of the FIM is not possible and so the CRLB cannot be found (we do not consider the case where $\Nxtavg{t_{\smM}}{ob} \to 0$, because we would not expect any meaningful communication if no molecules are expected at the RX). It has been shown in \cite{RefWorks:866,RefWorks:853} that if the FIM is singular, then there is no unbiased estimator for $\vectm{\theta}$ with finite variance. Furthermore, we must also consider the conditioning of the FIM. The $G_{\theta_i}$ terms associated with different parameters can vary by many orders of magnitude, such that the FIM can be nearly singular.

\section{Estimation Protocols}
\label{sec_protocol}

In this section, we describe the implementation of estimation protocols for the channel parameter estimation problem. First, we apply examples of ML estimation, as defined in Section~\ref{sec_ml_def}. We consider cases where the ML estimate can be written in analytical closed form. We also consider examples of applying the Newton-Raphson method to find the ML estimate numerically, and comment on comparing ML estimates with the CRLB when the FIM is singular or nearly singular.
Then, we propose peak-based estimation protocols as low-complexity methods for finding any one unknown channel parameter.

\subsection{ML Estimation}
\label{sec_ml}

\subsubsection{Analytical ML Estimation}
\label{sec_ml_anal}

We can try to search for ML estimates analytically by taking the derivative of the logarithm of the joint PDF with respect to the parameter of interest, i.e., (\ref{EQ14_04_156_gen}), and setting it equal to 0. If $\numParam > 1$, i.e., if there is more than one unknown parameter, then we will have to solve a system of equations (each in the form of (\ref{EQ14_04_156_gen})) to find the critical points that are candidates for the ML estimate. For tractability, we limit our discussion of analytical solutions to the special case of $\numParam = 1$ and $\M = 1$, and rely on numerical methods for the ML estimation of more than one parameter and/or observation. Furthermore, for ML estimation when $\numParam = 1$ and $\M = 1$, we use an approach that is more direct than taking the derivative of the logarithm of the joint PDF.

Consider the direct estimation of the expected channel impulse response at time $t_1$, $\Nxtavg{t_{1}}{ob}$. It is straightforward to show that the ML estimate of $\Nxtavg{t_{1}}{ob}$ is just the observation at time $t_1$, i.e., $\sx{1}$. Then, by the invariance property of ML estimation (see \cite[Ch.~3]{RefWorks:941}), the ML estimate of any single parameter in (\ref{EQ14_04_01}) can be found by setting $t = t_1$ in (\ref{EQ14_04_01}), substituting $\Nxtavg{t_{1}}{ob}$ with $\sx{1}$, and re-arranging to solve for the unknown parameter. Analytical solutions for estimating $t_0$ and $\Dx{}$ are not possible using this method because they are found both inside and outside the exponential in (\ref{EQ14_04_01}). We can still consider this method numerically for $t_0$ and $\Dx{}$ as an alternative to the numerical maximization of the likelihood function directly, except when $G_{\theta_i} = 0$.

The single-sample analytical ML estimates are then as follows:
\begin{align}
\label{ml_distance}
\hat{d}\textBar{ML} = &\; \vpara \ttext{ef} \pm
\sqrt{4\Dx{}\ttext{ef}H(\sx{1}) -\ttext{ef}^2(\vperp^2+4\kth{}\Dx{})},\\
\label{ml_degradation}
\hat{\kth{}}\textBar{ML} = &\; -\frac{\radmag{ef}^2}{4\Dx{}\ttext{ef}^2} + \frac{H(\sx{1})}{\ttext{ef}},\\
\label{ml_vpara}
\hat{\vpara}\textBar{ML} = &\; \frac{d}{\ttext{ef}} \pm \frac{1}{\ttext{ef}}
\sqrt{4\Dx{}\ttext{ef}H(\sx{1}) -\ttext{ef}^2(\vperp^2+4\kth{}\Dx{})},\\
\label{ml_vperp}
\hat{\vperp}\textBar{ML} = &\; \pm \frac{1}{\ttext{ef}}
\sqrt{4\Dx{}\ttext{ef}H(\sx{1}) -4\kth{}\Dx{}\ttext{ef}^2 - (d-\vpara \ttext{ef})^2},\\
\label{ml_Nemit}
\hat{\Nemit}\textBar{ML} = &\; \frac{\sx{1}(4\pi\Dx{}\ttext{ef})^{3/2}}{\Vrx}\EXP{\kth{}\ttext{ef} + \frac{\radmag{ef}^2}{4\Dx{} \ttext{ef}}},
\end{align}
where
\begin{equation}
H(\sx{1}) = \ln\left(\frac{\Nemit V_{RX}}{\sx{1}(4\pi\Dx{}\ttext{ef})^{3/2}}\right),
\end{equation}
we recall that $\radmag{ef}^2 = (d - \vpara \ttext{ef})^2 + (\vperp \ttext{ef})^2$, and here $\ttext{ef} = t_{1}-t_0$. Some additional comments on these ML estimates are necessary:
\begin{enumerate}
	\item $H(\sx{1})$ is a decreasing function of the observation $\sx{1}$. For a sufficiently large value of $\sx{1}$, an estimate of $\kth{}$ can be negative or an estimate of $d$, $\vpara$, or $\vperp$ can have an imaginary component. A negative degradation rate $\kth{}$ is physically meaningful and corresponds to the spontaneous generation of molecules in the propagation environment. Estimates with imaginary components should be ignored.
	\item The ``$\pm$'' in (\ref{ml_distance}), (\ref{ml_vpara}), and (\ref{ml_vperp}) mean that there could be multiple valid estimates due to the symmetry of (\ref{EQ14_04_01}) about the point $\{\vpara \ttext{ef}-d,0,0\}$.
	Even if the resulting distance $d$ is negative, it still has physical meaning because it represents uncertainty in the position of the TX relative to the RX, e.g., at $\{-d,0,0\}$ or $\{d,0,0\}$ if $\vpara = 0$. We could choose between multiple valid estimates by tossing an unbiased coin.
	\item If the observation $\sx{1} = 0$, then $H(\sx{1}) = \infty$ and all analytical ML estimates (except for that of $\Nemit$) are infinite. We can avoid infinite estimates by setting $\sx{1} = \sx{\epsilon}$ if $\sx{1} = 0$, where $0 < \sx{\epsilon} < 1$.
\end{enumerate}

As with the CRLB, we will find that the accuracy of ML estimation improves with the number of observations $\M$. Therefore, in Section~\ref{sec_num} we will not focus on assessing the above equations for single-sample analytical ML estimates.

\subsubsection{Iterative Numerical ML Estimation}

Here, we present examples of the structure of the Newton-Raphson method for numerically finding the ML parameter estimate. We consider the same examples that we examined in Section~\ref{sec_crlb_ex} because of their importance when establishing a communication link. First, we consider estimation of the distance $d$. Second, we consider the joint estimation of $d$ and $t_0$.

By (\ref{newton_raphson_def}), the distance $d$ can be found iteratively as
\begin{equation}
\label{newton_raphson_d_gen}
\hat{d}_{n+1} = \hat{d}_n - \pbypx{\ln p(\vect{\sx{}}|\vectm{\theta})}{d}\bigg/\psbypxs{\ln p(\vect{\sx{}}|\vectm{\theta})}{d} \bigg|_{d = \hat{d}_n},
\end{equation}
where we have already presented the first derivative of the logarithm of the joint PDF with respect to $d$ in (\ref{EQ14_04_156_gen}). The second derivative with respect to $d$ can be shown to be
\begin{equation}
\label{EQ14_04_128}
\psbypxs{\ln p(\vect{\sx{}}|\vectm{\theta})}{d} = 
-\sum_{\smM = 1}^\M \left(\frac{\sx{\smM}-\Nxtavg{t_{\smM}}{ob}}{2\Dx{}\ttext{ef}} + G_{d}^2\Nxtavg{t_{\smM}}{ob}\right),
\end{equation}
such that $d$ is found iteratively as
\begin{equation}
\label{newton_raphson_d}
\hat{d}_{n+1} = \hat{d}_n + \frac{\sum_{\smM = 1}^\M G_{d}\left(\sx{\smM} - \Nxtavg{t_{\smM}}{ob} \right)}{\sum_{\smM = 1}^\M \left(\frac{\sx{\smM}-\Nxtavg{t_{\smM}}{ob}}{2\Dx{}\ttext{ef}} + G_{d}^2\Nxtavg{t_{\smM}}{ob}\right)},
\end{equation}
where $G_{d}$ (as defined in (\ref{EQ14_04_156})) and $\Nxtavg{t_{\smM}}{ob}$ are evaluated for $d = \hat{d}_n$. Similar iterative expressions can be written for the iterative estimation of the other channel parameters. We see that, for a single observation, i.e., $\M = 1$, (\ref{newton_raphson_d}) will converge (such that $\hat{d}_{n+1}$ = $\hat{d}_n$) when the estimate $\hat{d}_n$ is such that $\sx{1} = \Nxtavg{t_{1}}{ob}$, unless we simultaneously have $G_{d} = 0$ (in which case the method will diverge).

The joint estimation of distance $d$ and synchronization (via $t_0$), such that $\vectm{\theta} = [d, t_0]^T$, can be found iteratively as
\begin{equation}
\label{newton_raphson_d_t_gen}
\begin{bmatrix}
\hat{d}_{n+1} \\ \hat{t}_{0_{n+1}}
\end{bmatrix} = 
\begin{bmatrix}
\hat{d}_{n} \\ \hat{t}_{0_{n}}
\end{bmatrix} - 
\begin{bmatrix}
\psbypxs{\ln p(\vect{\sx{}}|\vectm{\theta})}{d} & \psbypxpy{\ln p(\vect{\sx{}}|\vectm{\theta})}{d}{t_0} \\
\psbypxpy{\ln p(\vect{\sx{}}|\vectm{\theta})}{d}{t_0} & \psbypxs{\ln p(\vect{\sx{}}|\vectm{\theta})}{t_0}
\end{bmatrix}^{-1}\!
\begin{bmatrix}
\pbypx{\ln p(\vect{\sx{}}|\vectm{\theta})}{d} \\ \pbypx{\ln p(\vect{\sx{}}|\vectm{\theta})}{t_0}
\end{bmatrix},
\end{equation}
where we use $\hat{\vectm{\theta}}_n = [\hat{d}_n, \hat{t}_{0_{n}}]^T$ when we evaluate the derivatives of the logarithm of the joint PDF. It can be shown that the second derivative of the logarithm of the joint PDF with respect to $t_0$ is
\ifOneCol
\begin{equation}
\label{EQ14_04_129}
\psbypxs{\ln p(\vect{\sx{}}|\vectm{\theta})}{t_0} = 
\sum_{\smM = 1}^\M \left[\frac{\sx{\smM}-\Nxtavg{t_{\smM}}{ob}}{2\ttext{ef}^2}\left(3 - \frac{d^2}{\Dx{}\ttext{ef}}\right) - G_{t_0}^2\Nxtavg{t_{\smM}}{ob}\right],
\end{equation}
\else
\begin{align}
\label{EQ14_04_129}
\psbypxs{\ln p(\vect{\sx{}}|\vectm{\theta})}{t_0} = &\;
\sum_{\smM = 1}^\M \Bigg[\frac{\sx{\smM}-\Nxtavg{t_{\smM}}{ob}}{2\ttext{ef}^2}\left(3 - \frac{d^2}{\Dx{}\ttext{ef}}\right) \nonumber \\
&\;- G_{t_0}^2\Nxtavg{t_{\smM}}{ob}\Bigg],
\end{align}
\fi
and the cross derivative is
\ifOneCol
\begin{equation}
\label{EQ14_04_135}
\psbypxpy{\ln p(\vect{\sx{}}|\vectm{\theta})}{d}{t_0} = 
\sum_{\smM = 1}^\M \left(\frac{d\left(\Nxtavg{t_{\smM}}{ob}-\sx{\smM}\right)}{2\Dx{}\ttext{ef}^2} - G_{d}G_{t_0}\Nxtavg{t_{\smM}}{ob}\right).
\end{equation}
\else
\begin{align}
\label{EQ14_04_135}
\psbypxpy{\ln p(\vect{\sx{}}|\vectm{\theta})}{d}{t_0} = &\;
\sum_{\smM = 1}^\M \Bigg(\frac{d\left(\Nxtavg{t_{\smM}}{ob}-\sx{\smM}\right)}{2\Dx{}\ttext{ef}^2} \nonumber \\
&\;- G_{d}G_{t_0}\Nxtavg{t_{\smM}}{ob}\Bigg).
\end{align}
\fi

The structure of the Newton-Raphson method can be similarly described for other estimation problems with more than one unknown parameter.

\subsubsection{ML Estimation and the CRLB}

We complete our discussion of ML estimation by commenting on the behavior of ML estimation when the FIM is singular and the CRLB does not exist. Consider the estimation of a single parameter $\theta$ from a single observation so that from (\ref{EQ14_04_107_gen}) the FIM is a single element with no summation. If $G_{\theta} = 0$, then $\mathbf{I}\left(\theta\right)=0$ and no unbiased estimator with finite error variance exists. We have observed that we cannot find an analytical ML estimate when estimating one parameter $\theta$ from a single observation when $G_{\theta} = 0$. However, an informative ML estimate still exists; performing a finite grid search and choosing the estimate that maximizes the observation's log likelihood will result in a finite mean square error. We will see an example of this in Section~\ref{sec_num}. The ML estimate is still informative because it is now \emph{biased} (we previously noted in Section~\ref{sec_ml_def} that we can only claim that ML estimation is efficient in the sense of the CRLB as $\M \to \infty$).

It is insufficient to limit this discussion to the case where $\mathbf{I}\left(\theta\right) = 0$. In fact, ML estimation is biased and better than the CRLB when $\mathbf{I}\left(\theta\right)$ is in the ``vicinity'' of 0, i.e., as $\mathbf{I}\left(\theta\right) \to 0$.
Even in the case of estimating multiple parameters from a ``small'' number of observations, the FIM could be singular or nearly singular (this becomes less likely as more observation are made). Again, ML estimation in such a scenario can be biased and better than the CRLB. More seriously, poor conditioning can also cause convergence problems when implementing the Newton-Raphson method for ML estimation.

Existing literature (see \cite{RefWorks:854,RefWorks:855}) has sought to define the ``neighborhood'' of a singularity to determine where the CRLB is not an actual lower bound for ML estimation. However, this is a non-trivial task that has only been studied for some specific problems. A detailed study to determine the parameter values for which the CRLB is not a lower bound on ML estimation is outside the scope of this work. 

One might question whether knowledge of the CRLB is meaningful if it is not always a lower bound on ML estimation. We believe that it is relevant to have the CRLB because we are ultimately interested in practical parameter estimation schemes. A practical estimator will be more effective if it collects many observations over time. FIMs that are singular (and therefore have no corresponding CRLB) or close to singular will be less common as more observations are made, as we will observe in Section~\ref{sec_num}. Furthermore, ML estimation becomes unbiased (such that the CRLB is valid) as more observations are made. Thus, we claim that the CRLB is a useful benchmark.

\subsection{Peak-Based Estimation}
\label{sec_peak_est}

The study of parameter estimation in this paper has focused thus far on optimal performance, i.e., we have asked what is the best possible performance of an unbiased estimator and what is the performance of the maximum likelihood approach. We do not expect to implement a ML estimator as part of a nanoscale device, even if there is only one unknown parameter to estimate. Rather, our intent is to establish theoretical limits that we can use to compare with simpler, more practical estimators. We propose peak-based estimation for finding any one unknown channel parameter. Peak-based estimation has been proposed for distance estimation in \cite{RefWorks:614,RefWorks:776} and was also considered in our work in \cite{RefWorks:808}. It has been shown to be a relatively simple and accurate method for measuring the distance. By simple, we mean that a peak-based estimator makes multiple observations but uses just one observation to calculate the estimate.

In our simplest variation, the RX measures the time $\ttext{max}$ when the peak number of molecules is observed and uses the value of $\ttext{max}$ to estimate the unknown parameter. For comparison, we consider more complex variations where the RX measures the peak number of observed molecules $\stext{max}$, and also where the RX measures both $\ttext{max}$ and $\stext{max}$.

\subsubsection{Finding the Peak Time}

For peak-based estimation we need the time, after molecules are released by the TX, when the maximum number of molecules is \emph{expected}, i.e., $\overline{\ttext{max}}$ given that $t_0 = 0$. We previously derived $\overline{\ttext{max}}$ for our system model as \cite[Eq. (4)]{RefWorks:808}
\begin{equation}
\label{EQ14_04_02}
\overline{\ttext{max}} = \left(-3+\sqrt{9+d^2\eta/\Dx{}}\right)/\eta,
\end{equation}
where
\begin{equation}
\label{EQ14_04_03}
\eta = (\vpara^2 + \vperp^2)/\Dx{} + 4\kth{} = |\vxvec{}|^2/\Dx{} + 4\kth{}.
\end{equation}

In the absence of flow and molecule degradation, i.e., if $\eta = 0$, then it can be shown that the peak number of molecules would be expected at the RX at time
\begin{equation}
\label{EQ14_04_04}
\overline{\ttext{max}} = d^2/(6\Dx{}).
\end{equation}

Peak-based estimation requires the RX to measure either the peak number of observed molecules $\stext{max}$ or the time $\ttext{max}$ when the peak number is observed. The simplest method for doing so is to keep track of the number of molecules observed over a ``sufficiently'' long period of time and then select (either the time or the value of) the peak observation. A more general method, originally proposed in \cite{RefWorks:776}, is for the RX to track the upper and lower \emph{envelopes} of the observations. The ``peak'' observation $\stext{max}$ is then the peak value of the \emph{mean} of the two envelopes. We implemented the envelope detector in \cite{RefWorks:808} using what we called a moving maximum filter and a moving minimum filter. Given an odd filter length $\winLength$, the $m$th filtered observation $\sx{\smM}'$ of the moving minimum filter is
\begin{equation}
\label{def_max_filter}
\sx{\smM}' = \min_{w \in \{\smM-\frac{\winLength-1}{2},\ldots,\smM+\frac{\winLength-1}{2} \}} \sx{w},
\end{equation}
and the moving maximum filter is analogously defined. We note that a filter length $\winLength=1$ is analogous to the simplest method of determining $\stext{max}$ or $\ttext{max}$. We also note that the maximum observation $\stext{max}$ will generally be greater than the \emph{expected} observation at the time when the maximum observation is expected, even when using the envelope detector. This is discussed in greater detail in \cite{RefWorks:776}. The estimators that follow in the remainder of this section can be implemented with any method of finding (the time or the value of) the peak observation.

\subsubsection{Estimation from Peak Time}

Our simplest variation of peak-based estimation is when the RX estimates a parameter using $\ttext{max}$ alone (and not $\stext{max}$). If $t_0$ is the unknown parameter, then we assume that the RX is able to calculate $\overline{\ttext{max}}$ from (\ref{EQ14_04_02}) or (\ref{EQ14_04_04}) as appropriate and measure the time when the peak number of molecules is observed. If the \emph{observed} peak time is $\ttext{max}$, then the RX can immediately estimate $t_0$ as
\begin{equation}
\label{EQ14_04_26}
\hat{t}_0\textBar{Peak} = \ttext{max} - \overline{\ttext{max}}.
\end{equation}

For clarity of exposition in the remainder of this section, we will assume that $t_0 = 0$ when it is known and that the RX has adjusted its timer accordingly. Other values of $t_0$ can be accommodated by replacing the observed $\ttext{max}$ with $\ttext{max}-t_0$.

Estimates for \emph{most} of the remaining parameters can be derived by re-arranging (\ref{EQ14_04_02}) or (\ref{EQ14_04_04}) as appropriate (the number of molecules released, $\Nemit$, does not appear in (\ref{EQ14_04_02}) or (\ref{EQ14_04_04}), so we cannot use this method to estimate $\Nemit$). Generally, if we have flow or molecule degradation, i.e., if $\eta \ne 0$, then the remaining peak-based estimators are
\begin{align}
\label{EQ_peak_distance_est}
\hat{d}\textBar{Peak} = &\;\sqrt{\Dx{}\ttext{max}\left(4\kth{}\ttext{max}+6\right) + {\ttext{max}}^2|\vxvec{}|^2}, \\
\label{EQ14_04_42}
\hat{\Dx{}}\textBar{Peak} = &\; \frac{d^2 - {\ttext{max}}^2|\vxvec{}|^2}{4{\ttext{max}}^2\kth{} + 6\ttext{max}}, \\
\label{EQ14_04_46}
\hat{\kth{}}\textBar{Peak} = &\; \frac{d^2 - 6\Dx{}\ttext{max} - {\ttext{max}}^2|\vxvec{}|^2}{4\Dx{}{\ttext{max}}^2}, \\
\label{EQ14_04_52}
\hat{|\vxvec{}|}\textBar{Peak} = &\; \sqrt{\frac{d^2 - 4\Dx{}\kth{}{\ttext{max}}^2 - 6\Dx{}\ttext{max}}{{\ttext{max}}^2}},
\end{align}
and the estimators (\ref{EQ_peak_distance_est}) and (\ref{EQ14_04_42}) for the distance and the diffusion coefficient, respectively, also apply in the absence of flow and molecule degradation, i.e., if $\eta = 0$. We note that a \emph{two-way} version of the distance estimator (\ref{EQ_peak_distance_est}) when $\eta = 0$ was originally proposed as the round-trip time from peak concentration protocol in \cite{RefWorks:614}. Given $\hat{|\vxvec{}|}$ by (\ref{EQ14_04_52}) and the knowledge of one flow component, we can estimate the unknown flow component using $|\vxvec{}| = \sqrt{\vpara^2 + \vperp^2}$.

\subsubsection{Estimation from Peak Observation}

Our remaining peak-based estimation protocols are adapted from single-sample ML estimation, given that we have the peak observation $\stext{max}$. Any such parameter estimate will \emph{not} be the ML estimate given \emph{all} of the observations that were assessed to identify $\stext{max}$, but will be the single-sample ML estimate for the largest observation. These protocols must be implemented numerically, except for special cases, and are considered as (potentially) more accurate alternatives to estimation from only the peak time $\ttext{max}$.

If the RX has knowledge of both $\stext{max}$ and $\ttext{max}$, then both of these can be substituted into (\ref{EQ14_04_93}) and the ML estimate can be found numerically ($\ttext{max}$ is substituted for $t_1$). We can alternatively apply one of the analytical closed-form ML estimates found in Section~\ref{sec_ml} if the corresponding $G_{\theta_i} \ne 0$.

If the RX has knowledge of $\stext{max}$ but not of $\ttext{max}$, then the corresponding formula for $\overline{\ttext{max}}$ (either (\ref{EQ14_04_02}) or (\ref{EQ14_04_04})) can be substituted for $t_1$ in (\ref{EQ14_04_93}) and the ML estimate can be found numerically. This approach was applied in the implementation of the envelope detector proposed for distance estimation when $\eta = 0$ in \cite{RefWorks:776}. The only analytical ML estimate in Section~\ref{sec_ml} that remains in closed-form for any $\eta$ without requiring a numerical evaluation is that of $\Nemit$ in (\ref{ml_Nemit}) because $\overline{\ttext{max}}$ is not a function of the number of released molecules.

\section{Numerical and Simulation Results}
\label{sec_num}

In this section, we present simulation results to assess the performance of the channel parameter estimation protocols discussed in this paper with respect to the corresponding CRLBs. Our simulations were executed in the microscopic stochastic framework that we presented in \cite{RefWorks:662}. The TX is implemented as a spherical source such that the released molecules are initially separated by at least $1\,\textnormal{nm}$. The molecules are not created at a common point because they cannot physically occupy the same space, and $1\,\textnormal{nm}$ is larger than the size of single atoms and on the order of the size of small organic molecules that might be suitable for signaling; see \cite[Ch.~2]{RefWorks:588}. Every molecule released by the TX is treated as an independent point particle whose location is updated every simulation time step $\Delta t$. In one time step, a given molecule has a probability of $\kth{}\Delta t$ of degrading via reaction (\ref{k_mechanism}). All simulation results that we present in this section were averaged over $10^4$ independent simulations.

For clarity of exposition, since we have presented a number of parameter estimators in this paper, and there are many possible combinations of joint parameter estimation problems, we focus on a single set of environmental parameters as summarized in Table~\ref{table_param}. The RX has a radius of $0.5\,\mu\metre$, which is about the size of a small bacterial cell; see \cite[Ch. 1]{RefWorks:588}. The diffusion coefficient $\Dx{}$ of $10^{-9}\,\metre^2/\second$ is comparable to that of small molecules in blood plasma; see \cite{RefWorks:754}. The molecule degradation rate $\kth{}$ of $62.5\,\second^{-1}$ is sufficient, in the absence of flow, for an RX $4\,\mu\metre$ from the TX to expect one less molecule at the expected peak concentration time than if $\kth{} = 0$, i.e., $\Nxtavg{\ttext{max}}{ob} = 6.5$ instead of $7.5$.

\ifOneCol
\tableParam{!tb}
\else
\tableParam{!tb}
\fi

The flow magnitudes of $\vpara = 2\,\metre\metre/\second$ and $\vperp = 1\,\metre\metre/\second$ are strong relative to the diffusion but do not completely dominate; the Peclet number, which describes the relative dominance of convection versus diffusion and is found here as $d|\vxvec{}|/\Dx{}$ (see \cite[Ch. 5]{RefWorks:587}), is equal to $8.94$ when $d=4\,\mu\metre$. Such strong flows are within the range of average capillary blood speed (from $0.1$ to $10\, \metre\metre/\second$; see \cite{RefWorks:754}). We do not claim to accurately model capillary flow, where the flow is more complex than the uniform flow that we consider in this work, but such an environment is also one where the flow is relatively stronger than diffusion (without dominating; see \cite[Ch.~7]{RefWorks:750}). The strong flows also enable us to observe singularities in the CRLB for distance estimation at sampling times of interest (i.e., near when the maximum number of molecules is expected). The number of $\A$ molecules released by the TX at one time, $\Nemit = 10^5$, is the number of molecules that would be inside a spherical container of radius $0.5\,\mu\metre$ with a concentration of $0.32\,\metre\textnormal{M}$, which is at least an order of magnitude lower than the concentration of common ions used for signaling in mammalian cells; see \cite[Ch. 12]{RefWorks:588}.

Table~\ref{table_param} also lists the minimum and maximum parameter values that we use when performing a grid search of the maximum likelihood estimate of a given channel parameter. By symmetry, we only consider positive distance $d$ and positive perpendicular flow $\vperp$. We do not consider TX release times greater than $t_1$, the time of the first observation, because molecules cannot be observed before they are released. We also only consider non-negative degradation rate $\kth{}$, even though negative $\kth{}$ has physical meaning (i.e., information molecules are spontaneously created). Our constraints on the ranges of parameter values for grid searches enable ``best-case'' ML estimation; relaxing any of the constraints can only make ML estimation less accurate.

The resulting expected channel impulse response as a function of time, given the parameters listed in Table~\ref{table_param}, is presented in Fig.~\ref{fig_impulse} for varying distance $d$ from $2\,\mu\metre$ to $10\,\mu\metre$. We also show the average channel impulse response as generated by $10^4$ independent realizations of our simulator at each distance. Over this range of distances, the time of the expected maximum increases from about $\overline{\ttext{max}} = 0.5\,\metre\second$ to almost $\overline{\ttext{max}} = 4\,\metre\second$, and the number of molecules expected at that time decreases by almost two orders of magnitude. The average simulated responses are generally in agreement with the expected impulse responses, although the expected response tends to slightly underestimate the simulations before the peak time and overestimate the simulations after the peak time (due to the limitation of the assumption that the concentration expected throughout the RX is uniform). Assuming that the TX is a point source even though we simulate a spherical source is also a (negligible) source of inaccuracy.

\ifOneCol
\figImpulse{!tb}{0.7}
\else
\figImpulse{!tb}{1}
\fi

In the remainder of this section, we present normalized (i.e., dimensionless) results of the CRLB and the performance of the parameter estimators (unless otherwise noted). By normalizing our results, we are able to show the relative accuracy of estimating a given parameter. This is useful when a single parameter can vary over orders of magnitude, or when we want to show the estimation of different parameters on a single plot. We normalize the CRLB of parameter $\theta_i$ as
\begin{equation}
\label{EQ14_04_98_norm}
\frac{1}{\theta_{i_\textnormal{Ref}}^2}\left[\mathbf{I}^{-1}\left(\vectm{\theta}\right)\right]_{\theta_i,\theta_i},
\end{equation}
such that a CRLB of 1 means that the lower bound on the variance of an unbiased estimator is equal to $\theta_{i_\textnormal{Ref}}^2$. Generally, we will set $\theta_{i_\textnormal{Ref}} = \theta_i$. The one exception is that of $t_0$ because it has a value of $0\,\metre\second$. We set $t_{0_\textnormal{Ref}} = 0.1\,\metre\second$ so that the normalizing term is on the order of what an accurate estimate would be.

The performance of an estimator of parameter $\theta_i$ is evaluated by measuring the estimator's mean square error, which (unless otherwise noted) we normalize as
\begin{equation}
\label{mse_norm}
\mse{\hat{\theta}_i}\textBar{Norm} = E\left[(\hat{\theta}_i - \theta_i)^2\right]\big/\theta_{i_\textnormal{Ref}}^2,
\end{equation}
and we note that the non-normalized mean square error, i.e., without the scaling factor of $\theta_{i_\textnormal{Ref}}^2$, is equivalent to the variance in (\ref{var_def}) if and only if the estimator is unbiased. Generally, we aim for the CRLB and the mean square error to be as small as possible, such that the normalized bound and error should be much less than $1$ for the estimation to be meaningful.

Unless otherwise noted, the sampling scheme is as follows. When one observation is made, i.e., $\M=1$, then it is taken at $t_1 =2\,\metre\second$ (close to the time when the maximum number of molecules are expected at distance $d=6\,\mu\metre$; see Fig.~\ref{fig_impulse}). For other values of $\M$, the observation times are equally spaced such that the last sample is taken at time $t_\M =10\,\metre\second$. If we had only added new sample times when increasing $\M$ (without changing the old values of $t_\smM$), then from (\ref{EQ14_04_95}) and (\ref{EQ14_04_107_gen}) the CRLB could never increase. However, since we change the exact sample times for each value of $\M$, we will see results where the CRLB can increase with (small values of) increasing $\M$.

\subsection{Optimal Estimation}

We begin our discussion of estimator performance by focusing on optimal estimation, i.e., ML estimation and how it compares with the CRLB. All ML performance results presented were obtained via a grid search using the limits specified in Table~\ref{table_param}. By performing grid searches instead of using the analytical solutions available when $\M=1$, we do not need to address the exceptional cases described in Section~\ref{sec_ml_anal}. The performance of the estimation of a single parameter has also been verified via the Newton-Raphson method.

First, we consider estimating the distance when the true value is $d=6\,\mu\metre$ and we vary the number of observations made and the number of known parameters. We measure the normalized CRLB (given by (\ref{EQ14_04_98_norm})) and the normalized mean square error  (given by (\ref{mse_norm})) of ML estimation when only $d$ is unknown, and then successively remove the knowledge of $t_0$, $\vpara$, and $\vperp$. The results are shown in Fig.~\ref{fig_d_more_unknowns}. Removing the knowledge of $\Dx{}$, $\kth{}$, or $\Nemit$ is not as detrimental to distance estimation, so corresponding results are not shown. We will see later in this section that removing the knowledge of $d$ does not significantly degrade the estimation of $\Dx{}$, $\kth{}$, or $\Nemit$, either. We note that the ML estimate was solved for fewer values of $\M$ when there are three unknown parameters and for no values of $\M$ when there are four unknown parameters due to the increasing computational requirements of exhaustive searching. Applying the Newton-Raphson method for three and four unknown parameters was not feasible here due to poor matrix conditioning.

\ifOneCol
\figUnknowns{!tb}{0.7}
\else
\figUnknowns{!tb}{1}
\fi

In Fig.~\ref{fig_d_more_unknowns}, we see that there are no steady trends of ML estimation accuracy or its comparison with the CRLB for low values of $\M$, i.e., for $\M < 5$. This is for two reasons: the sampling times change significantly for each value of $\M$ and some of these samples are ``close'' to singular points. For example, a sample taken at about $t_{\smM}=3\,\metre\second$ will have a corresponding $G_{d}$ term with a value of $0$, which is why the CRLB when only $d$ is unknown and $\M=3$ (i.e., the first sample is at $t_1=3.3\,\metre\second$) is higher than when $\M=2$ (i.e., the first sample is at $t_1=5\,\metre\second$). We also cannot claim that losing knowledge of parameters will always degrade performance; when $\M=1$ or $5$, the ML estimate of $d$ when \emph{both} $d$ and $t_0$ are unknown is \emph{more accurate} than when \emph{only} $d$ is unknown. In these cases, the ML estimate trades accuracy in estimating $t_0$ for accuracy in estimating $d$ (later in this section, we will see that estimating $t_0$ is very inaccurate when $\M\le 5$). Nevertheless, we can make more general claims as more samples are taken, i.e., for $\M > 5$. As more samples are made, the CRLB improves and the ML estimate approaches the CRLB. In this regime, the CRLB and ML performance both degrade as more parameters become unknown. The potential mean square error in the estimation of $d$ increases by orders of magnitude as we remove the knowledge of the values of $t_0$, $\vpara$, and $\vperp$.

In Fig.~\ref{fig_single_all}, we observe the performance of ML estimation of each individual channel parameter when only that parameter is unknown. We set $d=6\,\mu\metre$, and we measure the normalized error of each parameter as a function of the number of samples $\M$. To ease inspection of the normalized error for small values of $\M$, this figure is shown in log-log scale. The figure gives us a sense of the relative accuracy to which we can aim to estimate any single channel parameter, and helps us to verify the diagonal elements of the FIM that we list in Table~\ref{table_fim}. As in Fig.~\ref{fig_d_more_unknowns}, the normalized error as a function of the number of samples begins to stabilize for $\M > 5$. We observe that the ML estimation of any single parameter performs very close to the corresponding CRLB as more samples are taken. In a relative sense, we can most accurately estimate the distance $d$, followed (in order) by the flow towards the RX $\vpara$, the perpendicular flow component $\vperp$, the number of molecules released $\Nemit$, the diffusion coefficient $\Dx{}$, the molecule degradation rate $\kth{}$, and finally the release time $t_0$ (although the choice of $t_{0_\textnormal{Ref}}$ was particularly arbitrary since we could not choose $\theta_{i_\textnormal{Ref}} = \theta_i$; the normalized error in the estimation of $t_0$ is comparable to that of $\vperp$ if we choose $t_{0_\textnormal{Ref}}=1\,\metre\second$ instead of $t_{0_\textnormal{Ref}}=0.1\,\metre\second$).

\ifOneCol
\figSingleAllML{!tb}{0.65}
\else
\figSingleAllML{!tb}{1}
\fi

In Fig.~\ref{fig_all_d_unknown}, we observe the performance of ML estimation of each individual channel parameter when there are two unknown parameters: the distance $d$ (whose actual value is still $6\,\mu\metre$) and the parameter of interest. We measure the normalized error of each parameter as a function of the number of samples $\M$. We can compare Fig.~\ref{fig_all_d_unknown} directly with Fig.~\ref{fig_single_all} to see the importance of the knowledge of the distance when estimating the other channel parameters. There is negligible degradation in the estimation of $\Dx{}$, $\kth{}$, and $\vperp$, slight degradation in the estimation of $\Nemit$, and significant degradation in the estimation of $t_0$ and $\vpara$. The negligible change in the estimation of $\vperp$ is most interesting because the opposite was not observed in Fig.~\ref{fig_d_more_unknowns}, where removing the knowledge of $\vperp$ was shown to measurably degrade the estimation of $d$.

\ifOneCol
\figAllDistUnknown{!tb}{0.65}
\else
\figAllDistUnknown{!tb}{1}
\fi

The results presented thus far do not give us a very clear sense of the performance of ML estimation in the ``neighborhood'' of a singularity in the FIM. To do so, we need to consider ML estimation as a function of a varying channel parameter whose domain includes a point where the CRLB is infinite. In Fig.~\ref{fig_d_singularity}, we perform distance estimation as a function of the actual distance $d$ for the number of observations $\M \in \{1,2,10,20,100\}$. We adjust the sampling times for $\M=2$ so that they are taken at $t_1=2\,\metre\second$ and $t_2=3\,\metre\second$. This adjustment ensures that, for every value of $\M$, a sample is taken at time $t_{\smM}=2\,\metre\second$, so that the corresponding $G_{d}$ is $0$ when $d=4\,\mu\metre$ (recall that $G_{d}$ is a function of $t_{\smM}$). Also, in this figure we do not normalize the mean square error or the CRLB because $d$ is the only unknown parameter.

\ifOneCol
\figSingularity{!tb}{0.65}
\else
\figSingularity{!tb}{1}
\fi

The only singularity in the FIM in Fig.~\ref{fig_d_singularity} is when $\M=1$ and $d=4\,\mu\metre$. Although ML estimation when $\M=1$ is generally not nearly as accurate as the CRLB, it is \emph{more} accurate than the CRLB over the range $3.2\,\mu\metre < d < 4.6\,\mu\metre$. This range is effectively the ``vicinity'' of the singularity when one sample is taken at time $t_1=2\,\metre\second$ and when ML estimation must be biased. Interestingly, when $\M=2$, there is never an actual singularity in the FIM, but we are still in the vicinity of a singularity when $3.6\,\mu\metre < d < 4.6\,\mu\metre$, where ML estimation is more accurate than the CRLB and must be unbiased. We observe this behavior where the $G_{d}$ term for the observation at $t_1=2\,\metre\second$ is equal to zero, but \emph{not} where the $G_{d}$ term for the observation at $t_2=3\,\metre\second$ is equal to zero, i.e., at around $d=6\,\mu\metre$. This is because the sample at time $t_1$ is more critical for the estimation of $d$ than that at $t_2$. The relative importance of individual samples is reduced as more samples are taken, such that ML estimation is only slightly more accurate than the CRLB at $d=2\,\mu\metre$ when $\M=10$, i.e., where the sample $t_1=1\,\metre\second$ is most critical for the estimation of $d$ and the corresponding $G_{d}$ term is $0$. Otherwise, we observe that ML estimation performs close to but not better than the CRLB for larger values of $\M$, where ML estimation becomes increasingly unbiased, over the entire range of $d$ that we consider.

\subsection{Peak-Based Estimation}

Finally, we consider the performance of the sub-optimal peak-based estimators that we proposed in Section~\ref{sec_peak_est}. We are interested in how well the simplest peak-based protocol (which only measures $\ttext{max}$ and can be implemented in closed-form) performs in comparison to the peak-based protocols that generally require a ML search given the value of the peak observation $\stext{max}$. The ML estimates given $\stext{max}$ are found via a grid search. We are also interested in the impact of the moving minimum (and maximum) filter window length $\winLength$ on the performance of each estimator, and whether the relative performance of the different estimators varies when different parameters are being estimated.

In Fig.~\ref{fig_peak_d}, we compare the performance of the peak-based distance estimators as a function of the actual distance $d$ for varying window length $\winLength$. Each estimator variation is described in a dedicated subplot. For reference and for comparison between subplots, we show the CRLB when a single sample is taken at time $t_1 = \overline{\ttext{max}}$ and when $\M=100$. Again, since $d$ is the only unknown parameter, we do not normalize the mean square error or the CRLB in this figure.

In all three subplots in Fig.~\ref{fig_peak_d}, no window length $\winLength$ emerges as optimal for the entire range of $d$. This makes sense and is consistent with our analysis of the envelope detector in \cite{RefWorks:808}; the best window length for a given distance is proportional to the time required for the diffusion wave to rise and then fall. Therefore, shorter filter lengths are more appropriate at shorter distances and longer filter lengths are generally better at longer distances.

More interestingly, the estimator that uses the knowledge of \emph{both} $\ttext{max}$ and $\stext{max}$ is much less accurate for measuring the distance than the estimators that use the knowledge of \emph{only} $\ttext{max}$ or $\stext{max}$. The reason is that this estimator uses the \emph{values} of $\ttext{max}$ and $\stext{max}$ but \emph{not} the knowledge that they correspond to the peak observation, i.e., neither (\ref{EQ14_04_02}) nor (\ref{EQ14_04_04}) are applied. Thus, the estimator does not ``know'' that its observation was made at time $\ttext{max}$. The ``simpler'' protocols combine the knowledge of $\ttext{max}$ or $\stext{max}$ with the knowledge that the observation was made at the peak time and use (\ref{EQ14_04_02}) or (\ref{EQ14_04_04}) as needed (i.e., depending on the value of $\eta$). The simplest protocol (using $\ttext{max}$) performs on the order of the single-sample CRLB for all window lengths over most distances, the protocol that uses only $\stext{max}$ often performs better than the single-sample CRLB, and the protocol using both $\ttext{max}$ and $\stext{max}$ always performs much worse than the single-sample CRLB.

\ifOneCol
\figPeakEstDist{!tb}{0.5}
\else
\figPeakEstDist{!tb}{1}
\fi

In Fig.~\ref{fig_peak_all}, we compare the performance of peak-based estimation of the other channel parameters as a function of the distance $d$. For clarity, we only consider a single window length $\winLength=7$. Each parameter is considered in a dedicated subplot. Where relevant, we show the CRLB when a single sample is taken at time $t_1 = \overline{\ttext{max}}$ (which is not applicable for $t_0$ because $G_{t_0}=0$ at that time) and when $\M=100$. The vertical scales here are not as important as the comparison between estimators and their performance relative to the CRLBs. Interestingly, the performance of the estimation of each parameter is not analogous to that of estimating the distance in Fig.~\ref{fig_peak_d}, which should not be too surprising because the peak-based estimators are sub-optimal \emph{ad hoc} methods. Instead, different peak-based estimators are more accurate at measuring different parameters. This is an important point when assessing the suitability of these estimation strategies. For example, the simplest protocol is the most accurate for estimating $t_0$ at shorter distances, but it is generally the least accurate when estimating $\kth{}$ or $\vperp$ at any distance. There is no clear best peak-based estimator for estimating $\Dx{}$ or $\Nemit$, whereas the estimator that uses only $\stext{max}$ is significantly more accurate than the other variants when estimating $\vpara$. Overall, the simplest estimator does not perform as well as the single-sample CRLB (when it exists) when estimating any parameter besides the distance, but both of the ML-based variants can perform better than the single-sample CRLB for some parameters.

\ifOneCol
\figPeakEstAll{!tb}{0.55}
\else
\figPeakEstAll{!tb}{1}
\fi

\section{Conclusion}
\label{sec_concl}

In this paper, we studied the local estimation of channel parameters when a transmitter releases impulses of molecules into a diffusive MC environment and the molecules are observed by a receiver. We considered an unbounded 3-dimensional environment with steady uniform flow and stochastic molecule degradation. We derived the FIM of the joint estimation problem, which leads to the CRLB on the error variance of any locally unbiased estimator. The FIM reduces for the estimation of any subset of the channel parameters. We considered ML estimation and presented cases where ML estimates can be evaluated in closed form. Generally, ML estimation is no more accurate than the CRLB, unless we are in the ``neighborhood'' of singularities in the corresponding FIM, but the impact of a sample being at or near a singularity diminishes as more samples are used in estimation. We proposed variations of peak-based estimation for more practical estimation of individual channel parameters, which rely on observing either the value or the time of the maximum number of molecules observed at the receiver.

The analysis presented in this work provides a benchmark for the future design of parameter estimation protocols. We are interested in the design of low-complexity estimators that use multiple samples (i.e., $\M > 1$) for estimation in more realistic environments. Low-complexity protocols would be more feasible in practice, but bounds on the accuracy of estimation give us insight into how much is lost by implementing sub-optimal solutions. Other related and interesting problems include cooperative estimation, where multiple devices share information to generate a common estimate, and channel estimation, where the expression for the expected channel impulse response is unknown and must be measured. Channel estimation is a more general problem because it does not rely on the existence of a closed-form expression for the expected impulse response.

\bibliography{../references/nano_ref}


\begin{IEEEbiography}[{\includegraphics[width=1in,height=1.25in,
		clip,keepaspectratio]{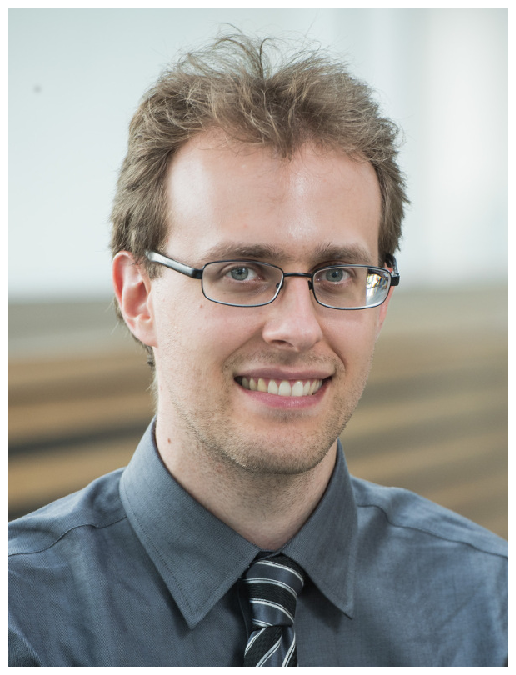}}]{Adam Noel}
	(S'09) received the B.Eng. degree from Memorial University in 2009 and the
	M.A.Sc.
	degree from the University of British Columbia (UBC) in 2011, both in
	electrical engineering.
	He is now a Ph.D. candidate in electrical
	engineering at UBC, and in 2013 was a visiting researcher
	at the Institute for Digital Communications,
	Friedrich-Alexander-Universit\"{a}t Erlangen-N\"{u}rnberg. His research
	interests include wireless communications and how traditional communication
	theory applies to molecular communication.
\end{IEEEbiography}


\begin{IEEEbiography}[{\includegraphics[width=1in,height=1.25in,
		clip,keepaspectratio]{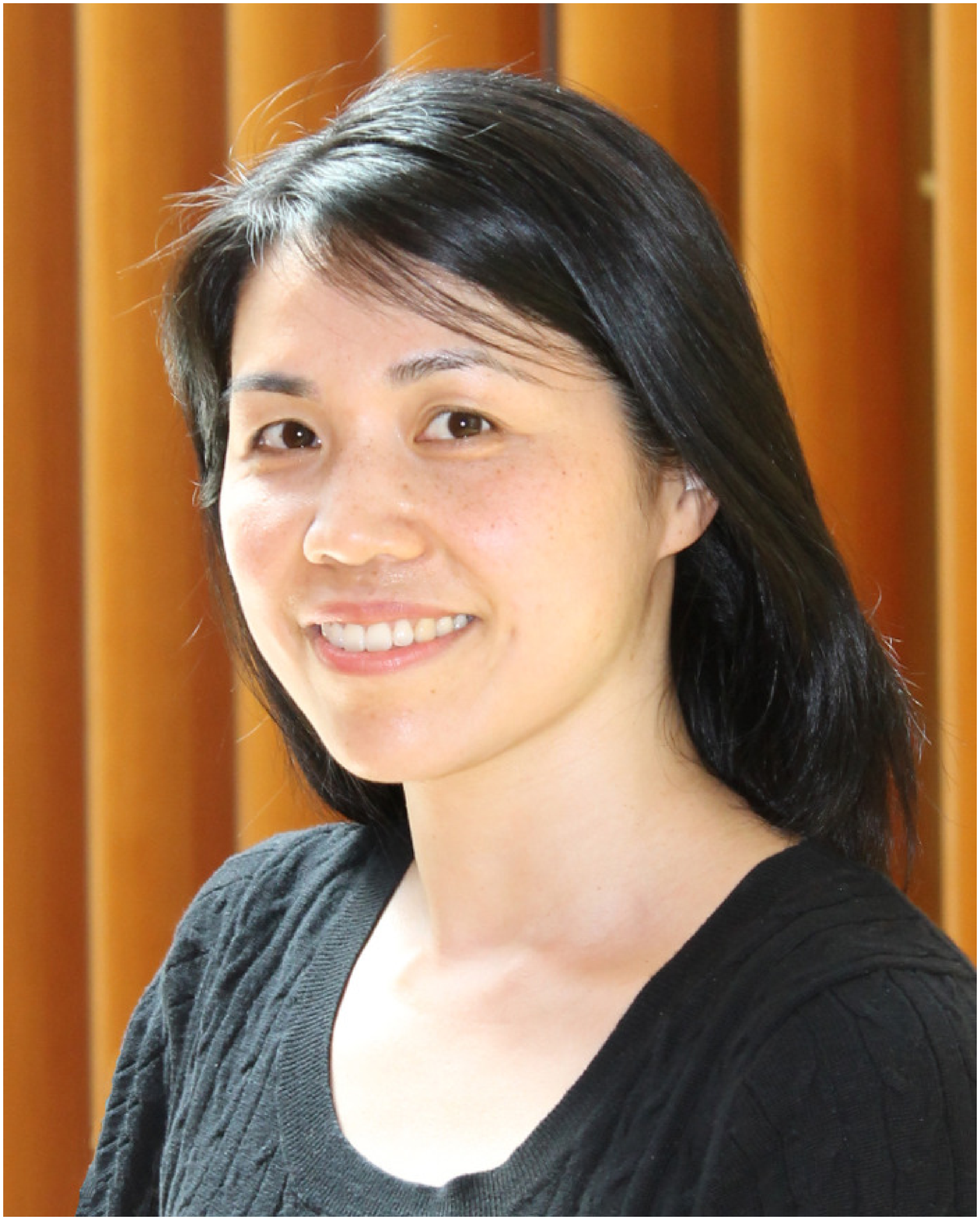}}]{Karen C. Cheung}
	received the B.S. and Ph.D. degrees in bioengineering from the University of
	California, Berkeley, in 1998 and 2002, respectively. From 2002 to 2005, she was
	a postdoctoral researcher at the Ecole Polytechnique Fédérale de Lausanne,
	Lausanne, Switzerland. She is now at the University of British Columbia,
	Vancouver, BC, Canada. Her research interests include lab-on-a-chip systems for
	cell culture and characterization, inkjet printing for tissue engineering, and
	implantable neural interfaces.
\end{IEEEbiography}


\begin{IEEEbiography}[{\includegraphics[width=1in,height=1.25in,
		clip,keepaspectratio]{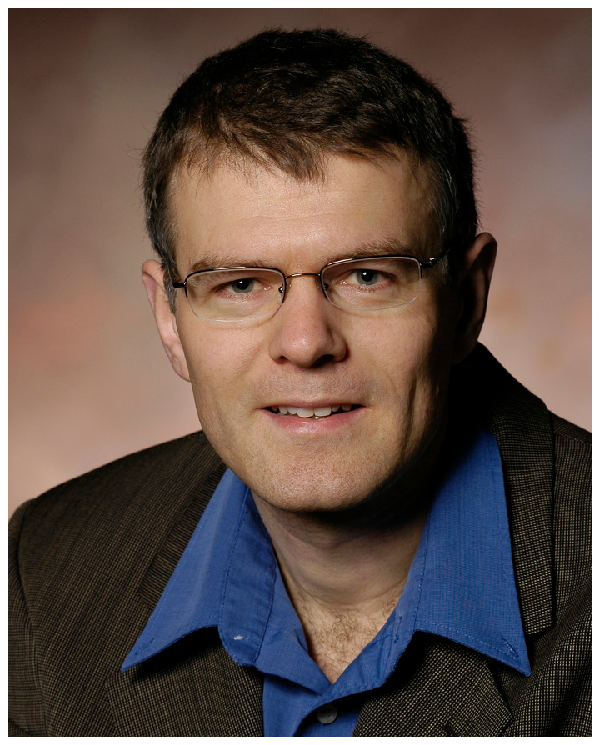}}]{Robert Schober}
	(S'98, M'01, SM'08, F'10)
	was born in Neuendettelsau, Germany, in 1971. He received the Diplom (Univ.) and the Ph.D. degrees in electrical engineering from the University of Erlangen-Nuermberg in 1997 and 2000, respectively. From May 2001 to April 2002 he was a Postdoctoral Fellow at the University of Toronto, Canada, sponsored by the German Academic Exchange Service (DAAD). Since May 2002 he has been with the University of British Columbia (UBC), Vancouver, Canada, where he is now a Full Professor. Since January 2012 he is an Alexander von Humboldt Professor and the Chair for Digital Communication at the Friedrich Alexander University (FAU), Erlangen, Germany. His research interests fall into the broad areas of Communication Theory, Wireless Communications, and Statistical Signal Processing.
	
	Dr. Schober received several awards for his work including the 2002 Heinz Maier–Leibnitz Award of the German Science Foundation (DFG), the 2004 Innovations Award of the Vodafone Foundation for Research in Mobile Communications, the 2006 UBC Killam Research Prize, the 2007 Wilhelm Friedrich Bessel Research Award of the Alexander von Humboldt Foundation, the 2008 Charles McDowell Award for Excellence in Research from UBC, a 2011 Alexander von Humboldt Professorship, and a 2012 NSERC E.W.R. Steacie Fellowship. In addition, he received best paper awards from the German Information Technology Society (ITG), the European Association for Signal, Speech and Image Processing (EURASIP), IEEE WCNC 2012, IEEE Globecom 2011, IEEE ICUWB 2006, the International Zurich Seminar on Broadband Communications, and European Wireless 2000. Dr. Schober is a Fellow of the Canadian Academy of Engineering and a Fellow of the Engineering Institute of Canada. He is currently the Editor-in-Chief of the IEEE Transactions on Communications.
	
\end{IEEEbiography}

\ifOneCol

	
		
	
	
	
	
	
	
	
	
\fi

\end{document}